\documentclass[
    amsmath, amssymb, aps,
    twocolumn, floatfix, showkeys,
    nofootinbib
]{revtex4-2}
\usepackage{graphicx}
\usepackage{dcolumn}
\usepackage{bm}
\usepackage{amsfonts}
\usepackage{physics2}
\usepackage{siunitx}
\usepackage{lineno}
\usepackage{makecell}
\usepackage{float}
\usepackage{color}
\usephysicsmodule{ab}
\usephysicsmodule{ab.braket}

\begin{document}

\title{Comprehensive Numerical Studies of Barren Plateau and Overparametrization \\ in Variational Quantum Algorithm}

\author{Himuro Hashimoto}
\email{Himuro.Hashimoto@yokogawa.com}
\author{Akio Nakabayashi}
\affiliation{%
 Yokogawa Electric Corporation, 2-9-32 Nakacho, Musashino-shi, Tokyo 180-8750, Japan
}

\author{Lento Nagano}
\email{lento@icepp.s.u-tokyo.ac.jp}
\author{Yutaro Iiyama}
\author{Ryu Sawada}
\author{Junichi Tanaka}
\author{Koji Terashi}
\affiliation{%
 International Center for Elementary Particle Physics,
 The University of Tokyo, 7-3-1 Hongo, Bunkyo-ku, Tokyo 113-0033, Japan
}

\date{February 3, 2026}

\begin{abstract}
The variational quantum algorithm (VQA) with a parametrized quantum circuit is widely applicable to near-term quantum computing, but its fundamental issues that limit optimization performance have been reported in the literature. 
For example, VQA optimization often suffers from vanishing gradients called barren plateau (BP) and the presence of local minima in the landscape of the cost function. Numerical studies have shown that the trap in local minima is significantly reduced when the circuit is overparametrized (OP), where the number of parameters exceeds a certain threshold.  Theoretical understanding of the BP and OP phenomena has advanced over the past years, however, 
comprehensive studies of both effects in the same setting are not fully covered in the literature.
In this paper, we perform a comprehensive numerical study in VQA, 
quantitatively evaluating the impacts of BP and OP and their interplay on the optimization of a variational quantum circuit, using concrete implementations of one-dimensional transverse and longitudinal field quantum Ising model. The numerical results are compared with the theoretical diagnostics of BP and OP phenomena. The framework presented in this paper will provide a guiding principle for designing VQA algorithms and ansatzes with theoretical support for behaviors of parameter optimization in practical settings.
\end{abstract}

\maketitle

\section{\label{sec:Introduction}Introduction}

The variational quantum algorithm (VQA) has received considerable attention in recent years, especially due to the potential for near-term application of quantum computing.
Typically, the VQA exploits a parametrized quantum circuit to transform a quantum state into a desired state for the problem of interest. In order to do this, the loss function is computed by a quantum computer, and the resulting loss function is processed via a classical optimizer to obtain optimal parameters of the quantum circuit.
While both the algorithm itself and the application have been extensively studied, it is well known that the optimization of the parametrized circuit, also referred to as the training, often suffers from fundamental problems, namely the phenomena of barren plateau (BP) and the presence of local minima.

The BP~\cite{McClean:2018jps} is a phenomenon where the variance of the gradient of the loss function vanishes exponentially with increasing number of qubits. Therefore, an exponentially large number of measurements is needed to navigate through the loss landscape based on the gradient, making the training cost prohibitively large. It is known that the BP can occur at various conditions depending on the expressibility of an ansatz~\cite{Holmes:2021qjw} and an encoding circuit~\cite{Thanasilp:2021axb}, entanglement powers of ansatz~\cite {Marrero:2020gvt,Patti:2020ach}, structure of observables~\cite{Cerezo:2020mtn}, and hardware noise~\cite{Wang:2020yjh}. 
In addition, for a certain choice of parametrized quantum circuits, the optimization process often exhibits local minima in the landscape of the VQA loss function. Once this occurs, the parameter update can be trapped at a faulty local minimum before reaching the global solution. 
It has numerically been observed that when the number of parameters exceeds a critical value, the loss function converges exponentially with the number of iterations~\cite{Wiersema:2020ipa,Kiani:2020bwb,Kim:2020luc}.
This phenomenon is called overparametrization (OP), and its theoretical foundation was established in~\cite{Larocca:2021jub}. Specifically, the authors gave a precise definition of OP by the quantum Fisher information matrix (QFIM) and related it to the dimension of the dynamical Lie algebra (DLA). 

\begin{figure*}[htbp]
\centering
\includegraphics[scale=0.9]{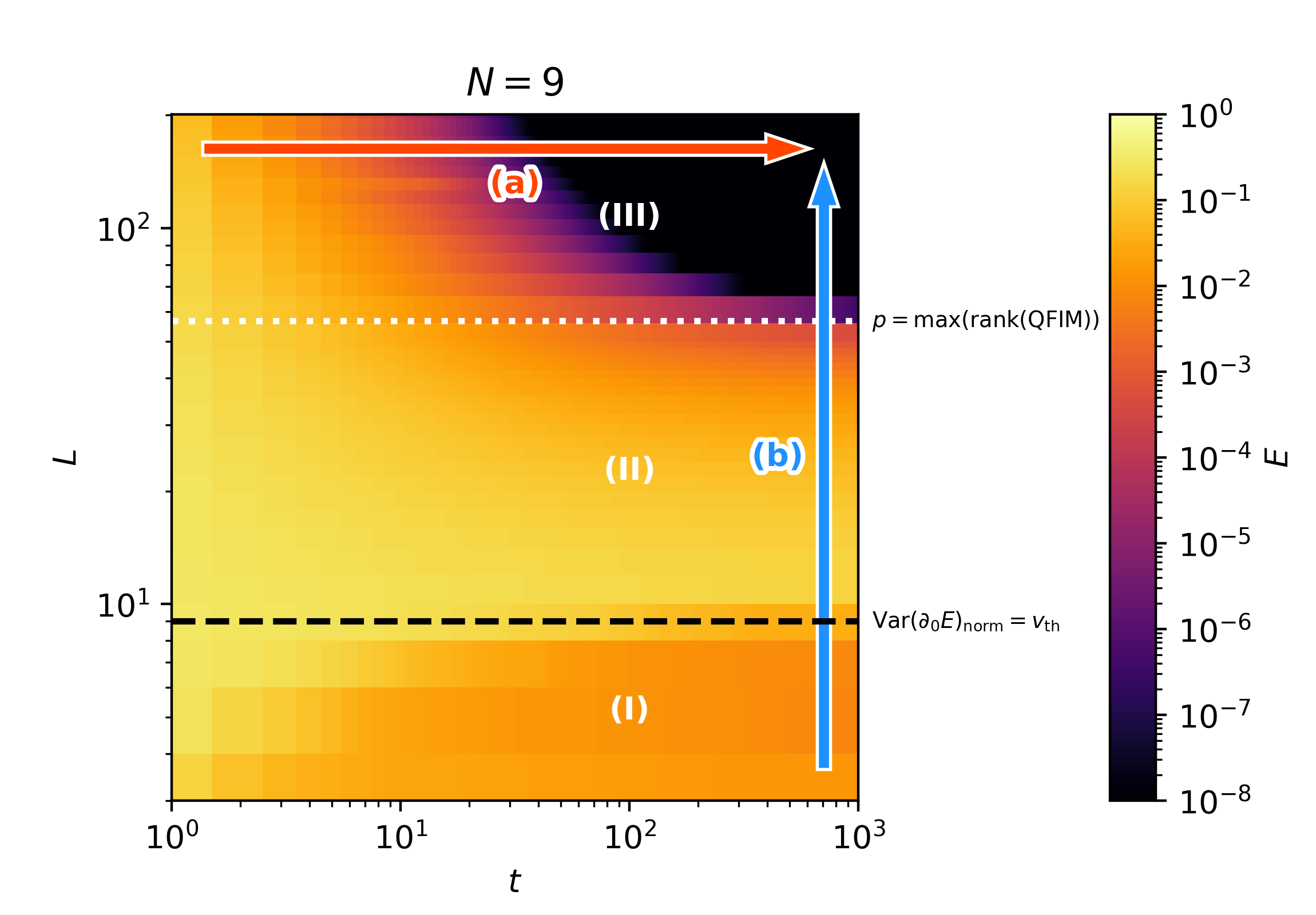}
\caption{Summary of numerical studies presented in this paper showing energy accuracy $E$ of VQE algorithm in the two-dimensional plane of the number of VQE ansatz layers ($L$) and the number of epochs ($t$). Detailed definitions of the presented quantities as well as the horizontal lines and the arrows are explained in the main text.}
\label{fig:intro-color-plot}
\end{figure*}

While BP and OP have been investigated thoroughly in the literature, most studies have focused only on each phenomenon separately. As illustrated by the arrows (a) or (b) in Fig.~\ref{fig:intro-color-plot}, this corresponds to the setting where the convergence of the VQA loss function is investigated only as a function of the training iterations or the number of parameters in the circuit, respectively.
Although the connection between the BP and OP via DLA has been theoretically understood, we still lack a unified framework to see both effects in a practical problem and quantify their impacts on the training loss. 
Such a framework will eventually allow us to choose appropriate ansatz structures, optimizers and hyperparameters (e.g., the number of parameters and layers in the circuit) for various applications.

In this paper, we perform a comprehensive numerical study of VQA performance by changing the number of parameters, iterations and the system size in practical problem settings. In particular, we focus on the Variational Quantum Eigensolver (VQE) as a representative example of VQA, and its convergence to the ground-state energy of a specific spin system using an ansatz tailored to the Hamiltonian.
In addition, we explicitly implement a variant of the sequential minimization in which the order of parameter updates is randomly shuffled at each epoch.
Our results can be summarized in Fig.~\ref{fig:intro-color-plot}, where the energy convergence in VQA is shown as a function of the number of layers in the circuit and the training iterations (details are discussed in the following sections).
In short, we observe three quantitatively distinct regions in the two-dimensional plane in terms of the energy accuracy (the value of the loss function), denoted by the regions (I), (II) and (III) in the figure.
For a sufficiently large number of iterations (at the slice (b) in Fig.~\ref{fig:intro-color-plot}), the boundaries of these regions can be understood from the maximal rank of the QFIM and the variance of the gradient of the loss function, both being consistent with {previous studies~\cite{McClean:2018jps,Larocca:2021jub}}. 
Moreover, the exponential convergence of the energy accuracy in the OP regime (at the slice (a) in the figure) agrees well with {the results in Ref.~\cite{Larocca:2021jub,Wiersema:2020ipa,Kiani:2020bwb,Kim:2020luc,Liu:2022eqa}}.
Interestingly, we not only reproduce previous findings, but also observe an intriguing effect in the region (III): the number of iterations necessary for converging the energy accuracy varies non-monotonically with the number of layers in the circuit.
This motivates further theoretical study to estimate how many parameters/iterations are required to observe a good convergence.

In order to understand the connection between the theoretical concept of learnability and the behavior in the practical implementation, we believe that a comprehensive study in practical settings with the visualization of training output, similar to the one presented in this paper, is useful for other benchmark ansatzes.
Moreover, this would help non-experts to learn how the circuit structure and choice of hyperparameters affect variational optimization, providing a guiding principle to design ansatz with theoretical support of the trainability.

\section{\label{sec:ExperimentalSettings}Problem setup}
VQE is a quantum-classical hybrid algorithm to find the eigenstates and eigenvalues of a given Hamiltonian~$H$.
A trial state called ansatz is first prepared via a parametrized unitary $U(\bm{\theta})$ as~$|\psi(\bm{\theta})\rangle=U(\bm{\theta})|\psi_0\rangle$ for an initial state $|\psi_0\rangle$, and then a cost function defined by $\langle\psi(\bm{\theta})|H|\psi(\bm{\theta})\rangle$ is evaluated from the measurement outcomes. The parameters are updated using a classical optimizer until the cost function converges.
In this section, we will explain our problem setup and notations used in this paper, which are summarized in Table~\ref{tab:experimental settings}.

\begin{table*}[tb]
    \centering
    \scalebox{0.9}{
        \begin{tabular}{|c|c|c|}
            \hline
            \textbf{Symbol} & \textbf{Definition} & \textbf{Setting in this paper} \\
            \hline
            $N$ & Number of qubits & {$3 \le N \le 10$} \\
            $L$ & Number of layers & 
            {$3 \le L \le 51$ in steps of $2$~($51 \le L \le 201$ in steps of $10$)}
            \\
            $t$ & Number of epochs & {$0 \le t \le 1000$} \\
            $r$ & Number of runs & {$r = 30$} \\
            $\mathcal{H}$ & Hamiltonian (density) & {TLFIM on a periodic chain $(J = h_X = h_Z = 1)$} \\
            $U(\bm{\theta})$ & Ansatz & {$\mathrm{HEA}$ composed from $(R_Y, R_Z)$ and CNOTs} \\
            $\bm{\theta}(t)$ & Parameters of ansatz & {$\bm{\theta}(t)\in[-\pi,\pi)^p$ randomly initialized at $t=0$} \\
            $p$ & Number of parameters & {$p=2NL$} \\
            $E(\bm{\theta}(t))$ & Relative residual energy & Eq.~\eqref{eq:residual-error}
            \\
            \hline
        \end{tabular}
    }
    \caption{Summary of experimental settings used in this paper.}
    \label{tab:experimental settings}
\end{table*}

\subsection{\label{sec:Hamiltonian}Hamiltonian}
\begin{figure}[tb]
    \centering
    \includegraphics[width=0.4\textwidth]{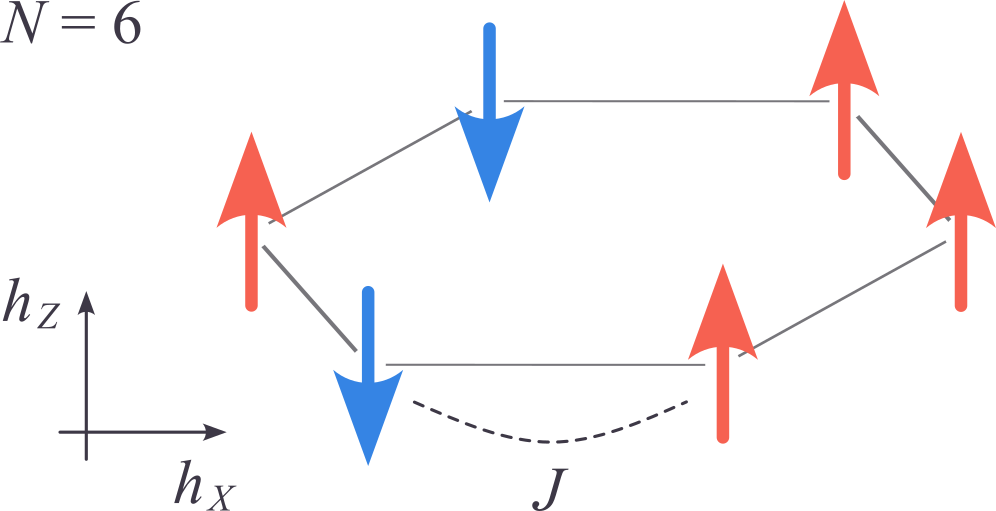}
    \caption{Schematic picture of the spin system for $N=6$ with the periodic boundary condition. The transverse and longitudinal fields $h_X$ and $h_Z$ are applied at each site, and each neighboring pair has an Ising interaction with strength~$J$.}
    \label{fig:TLFIM Hamiltonian}
\end{figure}

We consider the transverse and longitudinal field Ising model (TLFIM) Hamiltonian as a benchmark.
The Hamiltonian~\footnote{In this paper we consider the Hamiltonian normalized by the system size~$N$, which is usually called the Hamiltonian density.} is expressed by
\begin{align}
    \mathcal{H}_{\mathrm{TLFIM}} &= \frac{1}{N} \sum_{n=0}^{N-1} (J Z_{n}Z_{n+1} + h_X X_{n} + h_Z Z_{n})\,,
\end{align}
where
$N$ is the number of sites, and we assume the periodic boundary conditions, $P_{N} = P_{0}$, for~$\,P_i\in\{X_i,Y_i,Z_i\}$ acting on site $i$.
In the following experiment, we set~$J=1$, $h_X=1$ and $h_Z=1$.
A schematic picture of the spin-chain system is given in Fig.~\ref{fig:TLFIM Hamiltonian}.
The cost function is given by the energy measured in the ansatz state, that is,
the expectation value of the Hamiltonian (density) given as
\begin{align}
    \mathcal{E}(\bm{\theta})
    = \braket< 0 | U(\bm{\theta})^\dagger \mathcal{H} U(\bm{\theta}) | 0 >\,,
    \label{Eq:expval}
\end{align}
where $U(\bm{\theta})$ is a parametrized unitary that will be defined in the next subsection, and $\bm{\theta}$ is a set of parameters in the ansatz.
To quantify the accuracy of the measured energy with respect to the exact value, we use the following normalized difference between the VQE and exact results, defined as
\begin{align}\label{eq:residual-error}
    E(\bm{\theta})
    &:= \frac{\mathcal{E}(\bm{\theta}) - \lambda_{\min}}{\lambda_{\max} - \lambda_{\min}}\,,
\end{align}
where $\lambda_{\text{min}}$ ($\lambda_{\text{max}}$) denotes the minimal (maximal) eigenvalue of~$\mathcal{H}$ obtained by exact diagonalization, available for small systems.
The $E(\bm{\theta})$ value ranges from 0 (for the best case) to 1 (for the worst case). We will call this value the relative residual energy in the following.

\subsection{Ansatz} \label{sec:Ansatz}
The ansatz $U(\bm{\theta})$ can be chosen arbitrarily.
It should be sufficiently expressive to be able to represent a target eigenstate, while not too expressive to avoid barren plateau, as explained in Section~\ref{sec:BP}.
In this paper, we use the standard problem-agnostic ansatz, known as the Hardware Efficient Ansatz (HEA)~\cite{Kandala:2017vok}. 
One can systematically control its expressibility by changing the number of layers, making the ansatz a useful tool to investigate how the expressibility affects the trainability and accuracy.

\begin{figure}[tb]
    \centering
    \hspace{-0.2in}
    \includegraphics[scale=0.22]{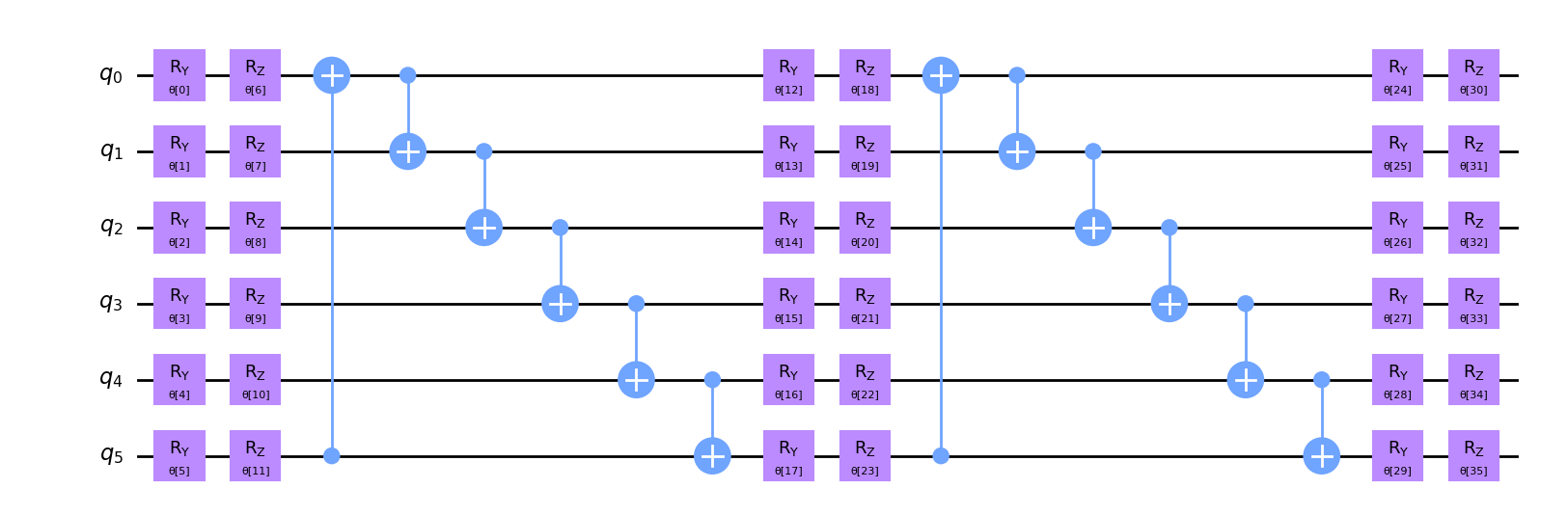}
    \caption{An example of HEA with $(N, L) = (6, 3)$ drawn using IBM \texttt{Qiskit} library~\cite{qiskit2024}. Each pair of parametrized block and entanglement block is repeated $L-1$ times, followed by another parametrized block.}
    \label{fig:ESU2 ansatz}
\end{figure}

The HEA consists of two types of blocks: parametrized blocks and entanglement blocks. In our setting, the parametrized blocks apply $R_Y$ and $R_Z$ gates sequentially to each qubit, while the entanglement blocks apply CNOT gates between neighboring pairs of qubits.
These blocks are repeated in the circuit, alternating between the parametrized blocks and entanglement blocks.
More precisely, it is defined as
\begin{align}
    U(\bm{\theta}) &= B_{L-1} \prod_{l=0}^{L-2} \ab(C \cdot B_l),\\
    B_{l} &= \prod_{n=0}^{N-1} R_{Z,n}(\theta_{2Nl+N+n}) R_{Y,n}(\theta_{2Nl+n}),\\
    C &= \prod_{n=0}^{N-1} {\mathrm{CNOT}}_{n,n+1},
\end{align}
where $N\ (\ge 2)$ is the number of qubits (identical to the number of sites in the TLFIM) and $L\ (\ge 2)$ is the number of parametrized blocks.
The total number of parameters is given by~$p = 2NL$.
An example of the HEA for~$(N, L) = (6, 3)$ is shown in Fig.~\ref{fig:ESU2 ansatz}.
In the numerical experiments in Section~\ref{sec:Results}, we vary $L$ from 3 to 51 in steps of 2, then from 51 to 201 in steps of 10, to examine the $L$ dependence.

\subsection{Optimization algorithm} \label{sec:Optimization algorithm}
After the ansatz is constructed,
the cost function is evaluated from the measurement outcomes and processed by a classical optimizer. A prototypical classical optimization is based on the gradient descent of the loss function derivatives.
In this paper, we use the sequential minimal optimization algorithm proposed by Nakanishi, Fujii and Todo, called NFT~\cite{Nakanishi:2019rrm}.
The NFT sequentially selects a single parameter to optimize at each step while fixing other parameters. Then, the optimization boils down to finding minima of a polynomial of trigonometric functions.
Each process of the single parameter optimization is referred to as \textit{step}, and a set of steps where all the parameters are scanned once as \textit{epoch} hereafter. 
In the standard NFT optimization, the order of sequential optimization is manually fixed, potentially leading to artificial influence on the convergence behaviors and the final accuracy.
To mitigate such effects, we introduce randomness in the order of parameter optimization, as indeed mentioned in the original NFT paper~\cite{Nakanishi:2019rrm} {and implemented, e.g., in~\cite{nicoli2023physics,masumoto_2025_15832836}}.
We randomly determine the order of parameters at each optimization step independently, referring to the NFT with this random ordering as epoch-wise random NFT (ERNFT) in the following.
We index steps as~$s\in\{1,2,\cdots, p\}$ and epochs as~$t\in\{0,1,\cdots,N_{\text{epochs}}\}$, respectively,
where $t=0$ represents the initial parameter configuration without any optimization.
In the first step of a given epoch $T$, $(s,t)=(1,T)$, we randomly select the parameter excluding the one in the final step of the previous epoch~$(s,t)=(p, T-1)$ 
to avoid redundant optimization steps. 
In the results section (Section~\ref{sec:Results}), the value of the cost function at each epoch~$t$ is taken from the value at the last step $s=p$.

In this paper, we mainly discuss the results obtained with ERNFT,
while performance comparison between the ERNFT and the standard gradient descent method is given in Appendix~\ref{sec:ERNFT vs others}.

\section{\label{sec:Results}Results}
In this section, we present the results of VQE optimization. To provide a comprehensive understanding, we analyze the data from multiple perspectives and visualize the results along with the reference quantities computed from generated quantum states and relative residual energy (i.e., QFIM rank and the gradient threshold). This helps us interpret the numerical results from a theoretical standpoint more easily.
 
\subsection{Experimental settings}
\label{sec:misc settings}
We perform the ERNFT optimization 30 times, each starting with different random initializations.
More precisely, the parameters of the $R_Y$ and $R_Z$ gates at the initial epoch $t = 0$ are chosen from a uniform random distribution for each run, such that $\theta_i \in [-\pi, \pi), i\in\{0,1,\cdots p-1\}$.
Unless otherwise noted, we present the average value of $E(\bm{\theta})$ over 30 runs at each epoch for each experimental setting.
All simulation results are obtained using the noiseless statevector simulator of the \texttt{Qulacs}~\cite{Suzuki:2020nbf} Python library.
The optimizers are also implemented with Python.

\begin{figure*}[tb]
    \centering
    \includegraphics[scale=0.7]{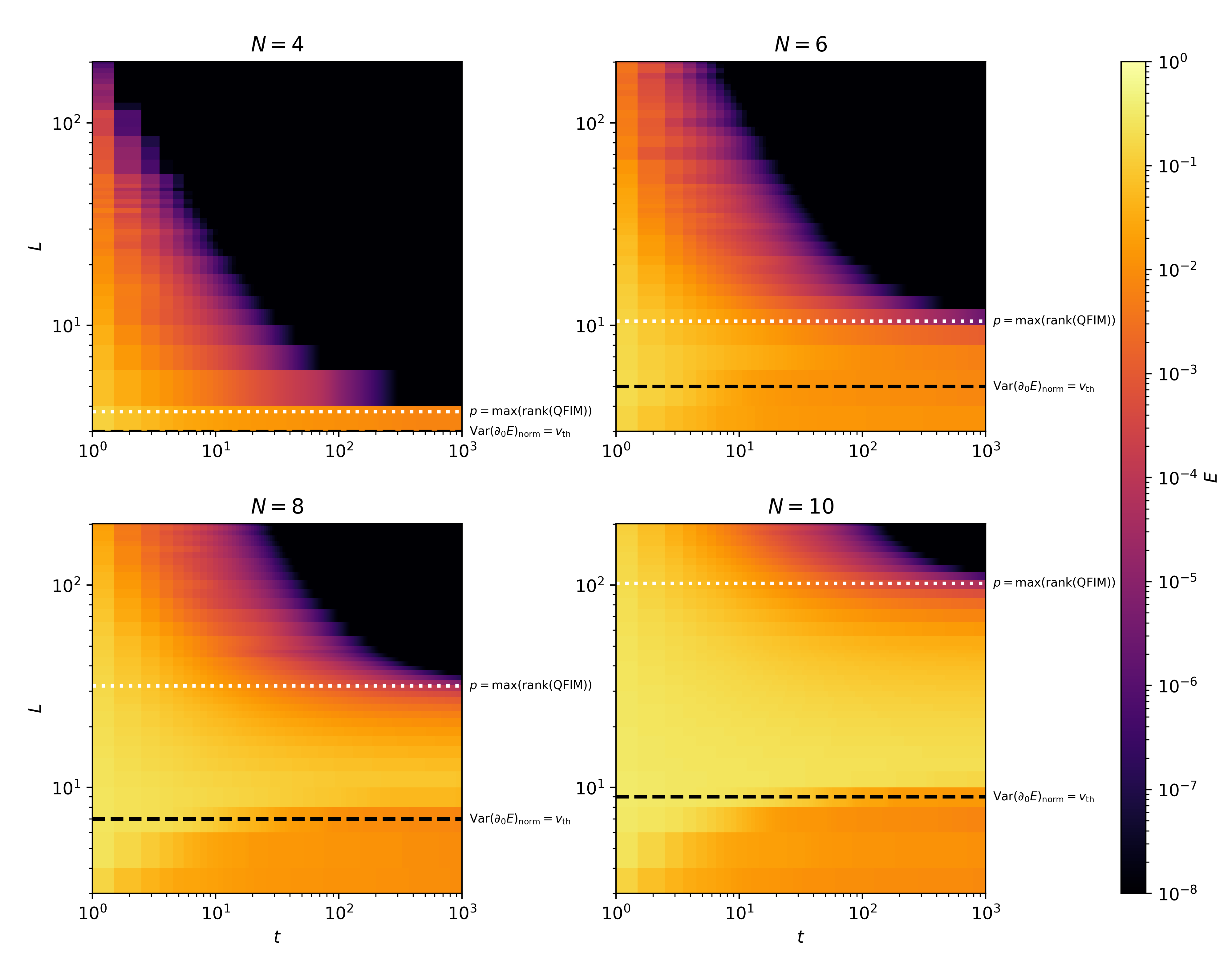}
    \caption{Heat maps of relative residual energy $E$ in the $t$-$L$ plane at $N=4$, 6, 8 and 10. As mentioned in Section \ref{sec:misc settings}, the $E$ value averaged over 30 runs is shown at each $(t, L)$ grid.
    The black dashed lines represent the locations where ${\mathrm{Var}(\partial_0 E)}_{\mathrm{norm}} = v_\mathrm{th}$, as discussed in Section \ref{sec:BP}.
    The white dotted lines correspond to the $L$ values where $p = \max(\mathrm{rank}(\mathrm{QFIM}))$, as discussed in Section \ref{sec:OP}.
    }
    \label{fig:tLEmap}
\end{figure*}

\subsection{Heat maps of optimization results}\label{sec:tLE}
First, we summarize the optimization results for $N\in\{4, 6, 8, 10\}$ in Fig.~\ref{fig:tLEmap}.
On top of that, the case for $N = 9$ is shown in Fig.~\ref{fig:intro-color-plot} as a representative example.
These figures show the dependence of relative residual energy $E(\bm{\theta})$ on the number of ansatz layers~$L$ and the number of epochs~$t$ at each system size~$N$. 
As seen in Fig.~\ref{fig:intro-color-plot}, we can observe three regions in each figure, distinguished by the values of relative residual energy, with the boundaries that coincide with the dotted and dashed lines at a sufficiently large value of~$t$.
Specifically, the white dotted line marks the point at which the number of parameters reaches the maximal rank of the QFIM, and the black dashed line is a boundary above which the normalized variance of the gradient is smaller than a certain threshold~$v_{\text{th}}$.
In the region below the black dashed line (region (I)), the gradient has a sufficiently large value for optimization, but the relative residual energy remains large due to poor expressibility of the ansatz.
In the region between the black and white lines (region (II)), the gradient becomes small, hence making the optimization harder. This explains larger 
relative residual energies in this region.
Once the value of~$L$ exceeds the white dotted line (region (III)), the ansatz is overparametrized and the redundant parameters help the optimizer find the optimal solution, making the residual energy converged to a very small value.
The gradient behavior and
the overparametrization are further discussed in Section~\ref{sec:BP} and~\ref{sec:OP}, respectively.

\subsection{Gradient of cost function and barren plateau}
\label{sec:BP}
In this section, we investigate the gradient of the cost function and discuss its effect on the optimization, known as barren plateau~\cite{McClean:2018jps, Larocca:2024plh}.

Barren plateaus in VQEs are defined by the variance of the gradients~$\partial_k E(\bm{\theta}):= \dfrac{\partial E(\bm{\theta})}{\partial \theta_k}$ that vanishes exponentially with increasing number of qubits.
Namely, 
the cost function is stated to exhibit a barren plateau when
\begin{align}
    \mathrm{Var}\ab( \partial_k E(\bm{\theta}) ) &:= \ab< (\partial_k E(\bm{\theta}))^2 > - \ab< \partial_k E(\bm{\theta}) >^2
    \notag
    \\
    &\propto \mathcal{O}(1/\alpha^N)\label{eq:BP-def-grad}
\end{align}
holds
for some $\alpha > 1$, where $\theta_k$ is a parameter chosen from a set of parameters in the ansatz.
This implies that the gradient of the cost function is close to zero over almost the entire parameter space, except for exponentially small regions, since its average value over the ansatz is zero
(assuming that it is constructed from Pauli rotations)~\footnote{
Another definition of barren plateaus involves a flat cost function landscape in most of the parameter space,
\begin{align}\label{eq:BP-def-loss}
    \mathrm{Var}(|E(\bm{\theta}_A) - E(\bm{\theta}_B)|)
    \propto \mathcal{O}(1/\beta^N),
\end{align}
where $\beta > 1$ is an arbitrary constant and $\bm{\theta}_A$ and $\bm{\theta}_B$ are parameter sets randomly sampled from the parameter space.
This definition is equivalent to the situation where Eq.~\eqref{eq:BP-def-grad} holds for all $k$, as shown in~\cite{miao2024equivalence,arrasmith2022equivalence}.
The behaviors of the left-hand side of Eq.~\eqref{eq:BP-def-loss} are shown in Appendix~\ref{sec:BP vardiff}.}.

\begin{figure}[tb]
    \centering
    \includegraphics[scale=0.75]{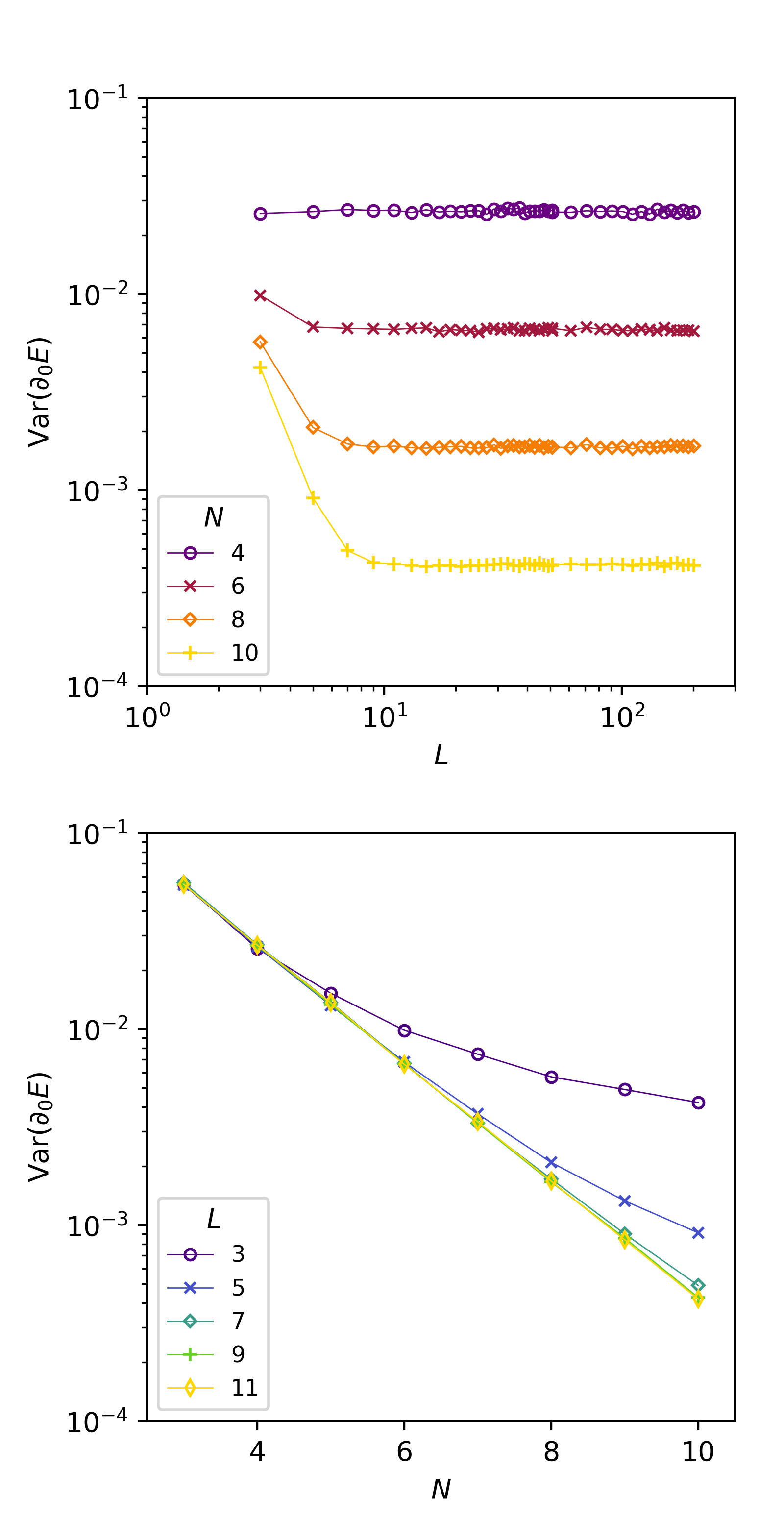}
    \caption{Variance $\mathrm{Var}(\partial_0 E)$ of partial gradients of relative residual energy with respect to the parameter $\theta_0$. The $\mathrm{Var}(\partial_0 E)$ is calculated from $10000$ random parameter sets at each point, and is shown as a function of $L$ ($N$) at fixed $N$ ($L$) values in the top (bottom) panel.}
    \label{fig:vargrad}
\end{figure}

Fig.~\ref{fig:vargrad} shows the variance of the gradient~$\mathrm{Var}(\partial_0 E)$ with varying $(N, L)$, where the parameter $\theta_0$ is the first parameter in the $R_Y$ gate in the first rotation block acting on the first qubit.
The variance of the gradients for each~$(N, L)$ is calculated from the gradients at $10^4$ randomly chosen points in the parameter space.
As shown in the figure, $\mathrm{Var}(\partial_0 E)$ vanishes exponentially as the number of qubits~$N$ increases for a sufficiently large number of layers~$L$, in agreement with previous works~\cite{Larocca:2021jub,Wiersema:2020ipa,Kiani:2020bwb,Kim:2020luc,Liu:2022eqa}.

To see the effects of a small gradient on the optimization, we define~$L^{\text{(BP)}}_{\text{th}}$ to be the minimal number of layers that satisfies 
\begin{align}
    \mathrm{Var}(\partial_0 E)_{\mathrm{norm}} :=\; \frac{\mathrm{Var}(\partial_0 E) - \mathrm{Var}(\partial_0 E)_\mathrm{min}}{\mathrm{Var}(\partial_0 E)_\mathrm{min}} < v_\mathrm{th}\,,
\end{align}
where~$\mathrm{Var}(\partial_0 E)_\mathrm{min}=\min_{L} (\mathrm{Var}(\partial_0 E))$.
We choose $v_\mathrm{th} = 0.05$ so that it can capture the convergence of the variance of gradients, and at the same time is sufficiently large compared to the statistical error of Monte Carlo sampling.
The values of~$L^{\text{(BP)}}_{\text{th}}$ are shown in Fig.~\ref{fig:tLEmap} as black dashed lines, and one can see that this is consistent with the boundary between the regions (I) and (II).
Above this boundary, the optimization becomes difficult and fails to find a good solution.
One would consider that barren plateau may have significant effects on the optimization.
On the other hand, the optimization can proceed easier in the region (I) below this line, indicated by lower relative residual energy in the region~(I) than the region~(II). The remaining residual energy in the region~(I) can be attributed to the poor expressibility of the ansatz.   
The expressibility can be quantified by the frame potential, which is discussed in Appendix~\ref{BP frame potential}.

\subsection{\label{sec:OP}QFIM and Overparametrization}

For a typical cost function expressed by Eq.~(\ref{Eq:expval}), the function landscape may contain many local minima, causing the parameter optimization to be trapped in one of those minima. When there are many parameters, it is known that these local minima often vanish, and the optimization becomes much easier. The cost function in this regime is referred to as overparametrized~\cite{Larocca:2021jub}.
Ref.~\cite{Larocca:2021jub} gave a precise definition of the OP using
QFIM~\cite{Petz:2010avh}.
First, the QFIM is defined as
\begin{equation}
\begin{split}
    \mathcal{M}_{ab}
    &= 4 \mathrm{Re}\big( \braket<\partial_a \psi(\bm{\theta}) | \partial_b \psi(\bm{\theta})> \\
    &\hspace{4em} - \braket<\partial_a \psi(\bm{\theta}) | \psi(\bm{\theta})> \braket<\psi(\bm{\theta}) | \partial_b \psi(\bm{\theta})> \big)\,,
\end{split}
\end{equation}
where $\ket|\partial_a \psi(\bm{\theta})> = \frac{\partial |\psi(\bm{\theta})\rangle}{\partial \theta_a}$
is computed with respect to the $a$-th parameter in $\bm{\theta}$.
Formally, the maximum rank of the QFIM is defined for a given $L$ as
\begin{equation}
    \mathcal{R}_L = \max_{\bm{\theta}}\left(\text{rank}\left(\mathcal{M}_{ab}\right)\right),
\end{equation}
and it saturates at a value given by  
\begin{equation}
    \mathcal{R}_{\text{max}} = \max_{L\in{\mathbb Z}} \left(\mathcal{R}_L\right).
\end{equation}
Therefore, we define the overparametrization threshold to be
\begin{equation}
    L^{\text{(OP)}}_{\text{th}} = \arg\min_{L}\left(\mathcal{R}_L\,|\,\mathcal{R}_L = \mathcal{R}_{\text{max}}\right),
\end{equation}
and the ansatz with $L>L^{\text{(OP)}}_{\text{th}}$ is overparametrized.
The QFIM rank is bounded by the number of independent coefficients to span the entire Hilbert space, which is 
\begin{align}\label{eq:rank-QFIM-controllable}
    \mathcal{D} = 2^{N+1}-2\,.
\end{align}
\begin{figure}[tb]
    \centering
    \includegraphics[scale=0.75]{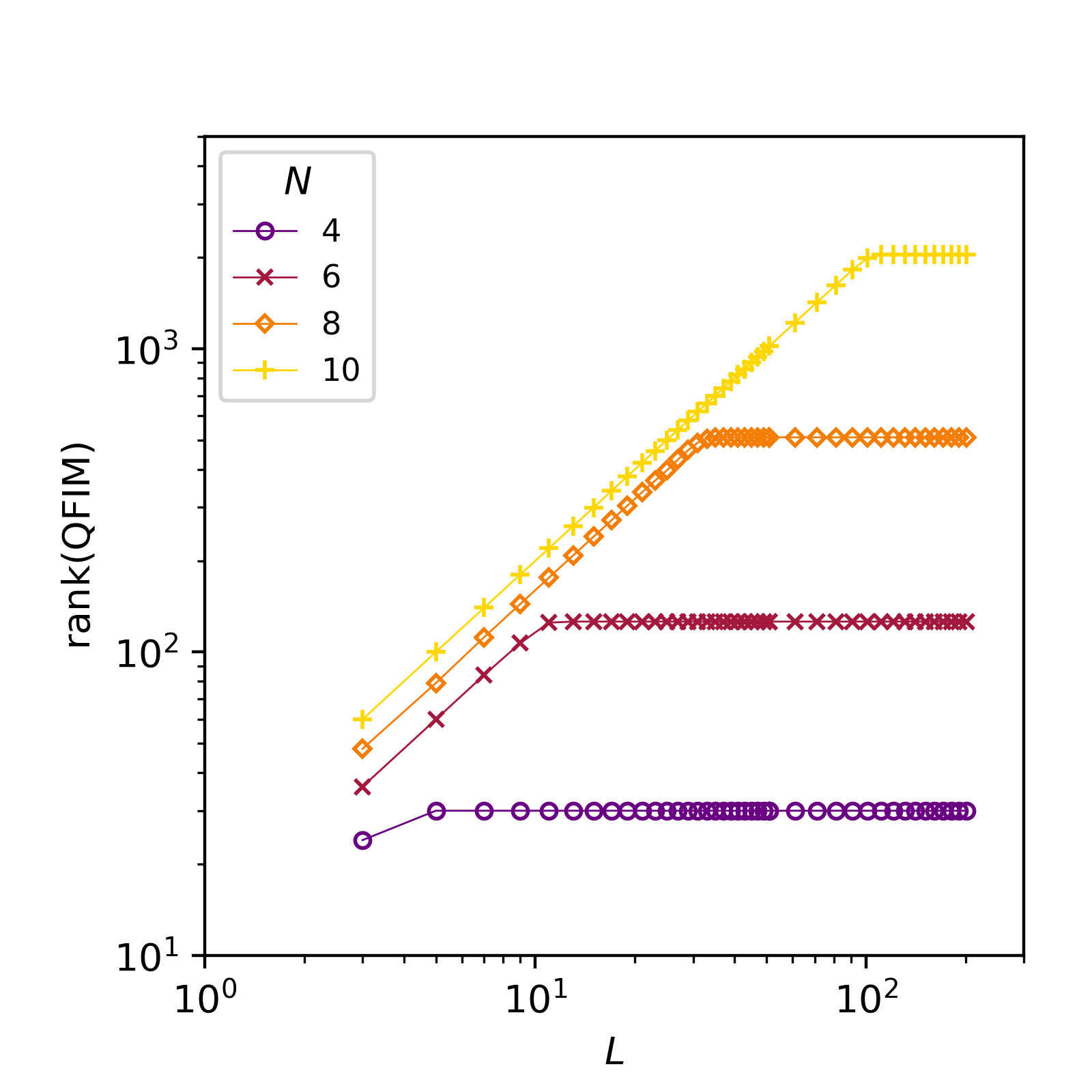}
    \caption{Rank of the QFIM as a function of $L$ at $N=4$, 6, 8 and 10.
    The calculated values are connected by a line at each $N$ for better visualization.
    The saturated value of the QFIM rank is given by $2^{N+1}-2$.}
    \label{fig:rankQFIM}
\end{figure}
We show the~$L$-dependence of the QFIM rank~$\mathcal{R}_L$
for each~$N$ in Fig.~\ref{fig:rankQFIM}. One can observe that the rank saturates at the value of $L=L^{\text{(OP)}}_{\text{th}}$ and the saturated value is given by Eq.~\eqref{eq:rank-QFIM-controllable}. The values of $L^{\text{(OP)}}_{\text{th}}$ obtained from this analysis are shown as the white dotted lines in Fig.~\ref{fig:tLEmap}, and agree well with the boundaries between the region~(II) and region~(III). 
This observation is consistent with the general understanding that the optimization becomes much easier after the QFIM rank is saturated (i.e., the ansatz in the overparametrized regime), due to the benign cost function landscape~\cite{Larocca:2021jub}. 

\begin{figure*}[tb]
    \centering
    \includegraphics[scale=0.7]{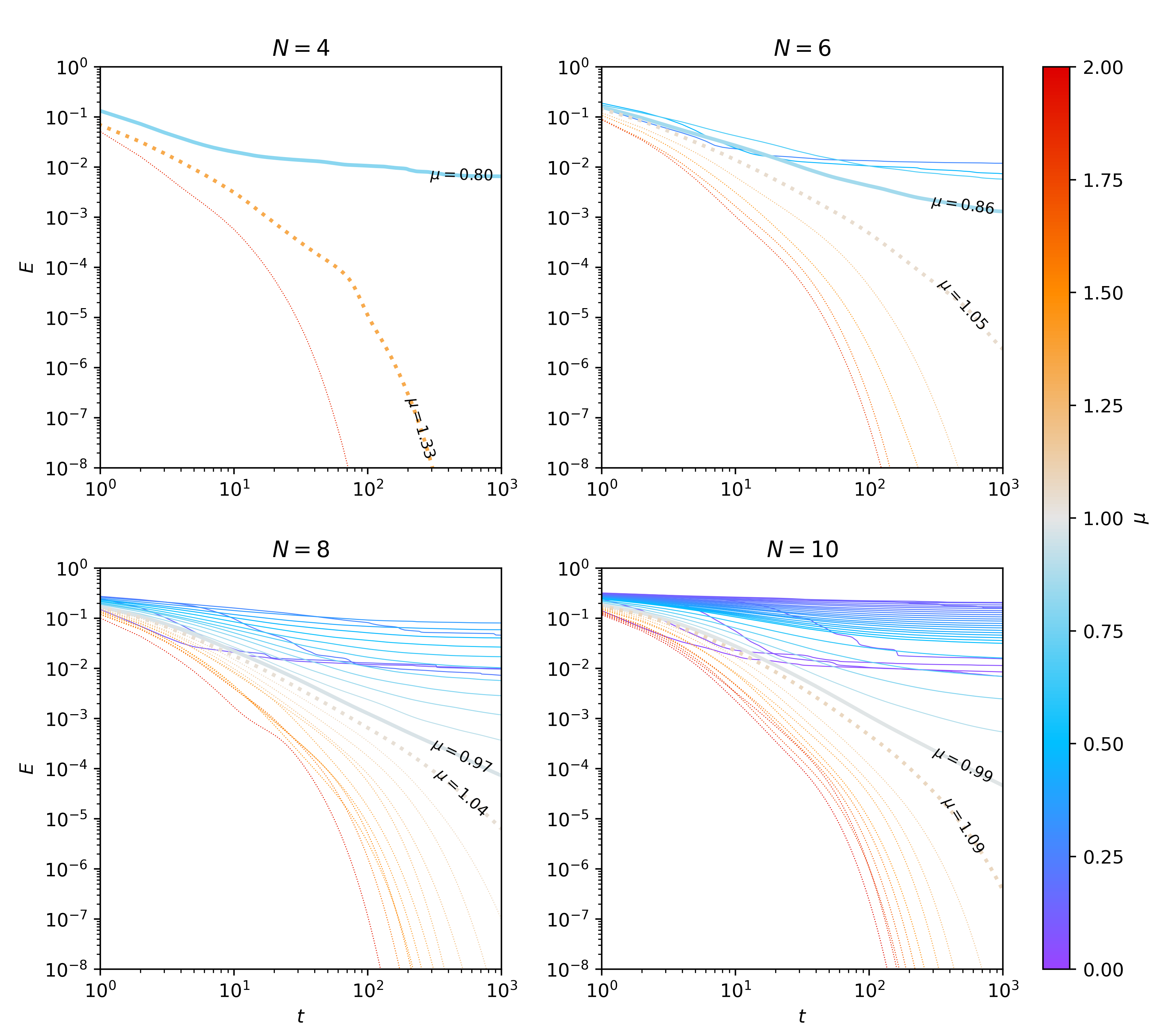}
    \caption{Relative residual energy $E$ as a function of $t$ at $N=4$, 6, 8 and 10. Each curve is obtained from the average over 30 runs. In each figure, the solid blue (dashed red) lines represent $E$ at $\mu < 1$ ($\mu > 1$). The residual energy for the $\mu$ value closest to 1 is shown by the gray thicker line for $N=8$ and 10.}
    \label{fig:tELN}
\end{figure*}

To visualize the convergence rate before and after the overparametrization, we also show how the relative residual energy~$E$ varies with~$t$~(iteration) at the fixed number of parameters~$p=2NL$ in~Fig.~\ref{fig:tELN}. This corresponds to the slice (a) shown in Fig.~\ref{fig:intro-color-plot} (but for different~$N$).
In this figure, the residual energies are averaged over 30 runs, and the number of parameters~$p$ is normalized as $\mu=p/(2^{N+1}-2)$.
Furthermore, the convergence curves with the number of parameters closest to the saturated value of $\mu=1$ (i.e., $L=L^{\text{(OP)}}_{\text{th}}$) among the scanned parameter range are shown by the gray lines for $N=8$ and 10, and the curves in the over(under)parametrized regime are shown by the blue (red) lines.
When the ansatz is very shallow at $\mu\lesssim 0.1$, the relative residual energy decreases at early epochs and is trapped around~$E\simeq 10^{-2}$. The curves become a smoothly falling function when the value of~$\mu$ approaches one, and finally show a polynomial decay of $E \propto \mathcal{O}\ab(t^\mathrm{const.})$ around $\mu \simeq 1$. To the best of our knowledge, this polynomial decay around~$\mu\simeq1$ has not been observed yet, and the theoretical understanding would be an open problem~\footnote{Power-law or sublinear behaviors due to specific choices of data label and eigenvalue of observables in a supervised quantum machine learning task were discussed in~\cite{Zhang:2024ktz,you2023analyzing}. However, the relation to the power-law behavior observed here is not clear.}.
Once the ansatz reaches the overparametrized regime of~$\mu>1$, we observe the exponential convergence of the residual energies $E \propto \mathcal{O}(\mathrm{e}^{-t})$, in agreement with previous works~\cite{Larocca:2021jub,Wiersema:2020ipa,Kiani:2020bwb,Kim:2020luc,Liu:2022eqa}.

\subsection{\label{sec:LossJump}$L$-dependence}
The transitions to the BP and OP regimes can be seen more clearly by looking at $L$-dependence of the relative residual energy $E$ with fixed~$N$ at a sufficiently large iteration of $t=10^3$, as shown in Fig.~\ref{fig:LENt}~\footnote{{The similar plots for $t=10$ and $100$ are shown in Appendix~\ref{sec:results appendix}. The qualitative behaviors are similar to the case of $t=10^3$, but the quantitative behaviors are different. In particular, the boundaries deviate from the theoretical reference points.}}.
This corresponds to the vertical slice (b) in Fig.~\ref{fig:intro-color-plot}.
In the top panel of this figure, we observe that for $N \gtrsim 8$, there is a sudden increase in $E$ with increasing $L$.
We refer to this sudden increase as a ``jump'' in $E$.
These jumps occur around $L \simeq 10$, followed by the decrease in $E$ at larger~$L$. 
This behavior can be understood as follows: the ansatz resides in the region with poor expressibility (denoted as region (I) in Fig.~\ref{fig:intro-color-plot}) before the jump occurs. Then, it moves to the BP region (region (II)) when the $L$ reaches around 10 and the jump occurs. Further increase of $L$ drives the ansatz into the OP region (region (III)), where the relative residual energy falls exponentially to a very small value.

The bottom panel in Fig.~\ref{fig:LENt} presents the same data as a heat map.
The white circles show the OP thresholds (corresponding to the white dotted lines in Fig.~\ref{fig:tLEmap}), and the black crosses show the BP thresholds (the black dashed lines in Fig.~\ref{fig:tLEmap}).
The regions (I), (II) and (III) are also visible in this figure, distinguished by the different $E$ values.
One can observe that the region (II) (the region between the circles and crosses), corresponding to the under-parametrized ansatz with a small gradient, becomes larger as the system size~$N$ increases. This implies that the optimization for the HEA does not scale efficiently with $N$, which is consistent with the DLA analysis~\cite{Larocca:2021ksf,Larocca:2021jub}.

\begin{figure}[tb]
    \centering
    \includegraphics[scale=0.7]{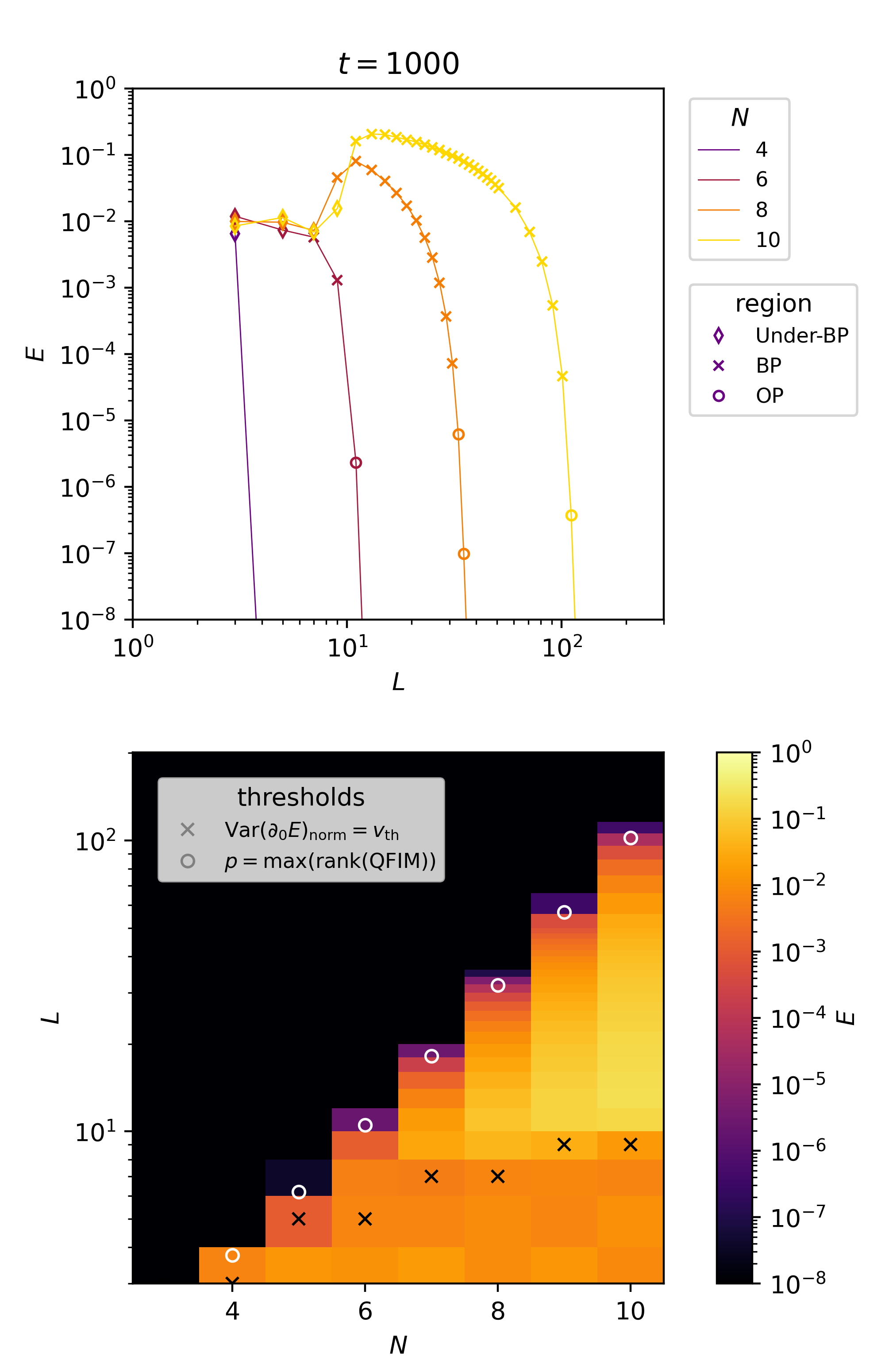}
    \caption{(Top panel) Relative residual energy $E$ as a function of $L$ at $N=4$, 6, 8 and 10 and $t=1000$.
    The markers show the averaged $E$ values calculated at each setting, and the lines connect the values for better visualization. (Bottom panel) Relative residual energy $E$ in $N$-$L$ plane with the white circles (black crosses) showing the OP (BP) thresholds, as explained in the text. }
    \label{fig:LENt}
\end{figure}

\section{\label{sec:Conclusion}Conclusion}
In this paper, we performed a comprehensive numerical study for a variational quantum algorithm in TLFIM to benchmark ansatz structures, optimizers and hyperparameters for the VQA trainability.
Specifically, we measured the ground-state energy of TLFIM using VQE with fixed coupling constants, and investigated the dependence of the energy accuracy on the number of layers~$L$, the number of iterations~$t$ and the number of qubits~$N$. 

Our main results are summarized in Fig.~\ref{fig:tLEmap}, showing the relative residual energy $E$ in the 2-dimensional plane of $L$ and~$t$. On top of that, the white dotted lines show the $L^{\text{(OP)}}_{\text{th}}$ obtained from the maximal rank of the QFIM, and the black dashed lines represent the boundaries above which the gradient of the cost function becomes smaller than the threshold. In a sufficiently large~$t$ region, we observe that the $L$-dependence is consistent with previous works: precisely, the residual energy gets larger when $L$ exceeds the boundary due to BP effects, and it becomes smaller again at $L$ above $L^{\text{(OP)}}_{\text{th}}$ due to overparametrization.
The behavior in the large~$L$ region can also be understood from previous studies. The horizontal scan in this region is shown in Fig.~\ref{fig:tELN}, and one can observe that the value of the cost function exponentially decreases at $L$ above $L^{\text{(OP)}}_{\text{th}}$. A theoretical understanding of the $L$-dependence of the decay rates is an important future direction. Besides, we have observed that the $E$ exhibits a power-law decay in $t$, appearing as a linear behavior in Fig.~\ref{fig:tELN}, near the QFIM boundary of $\mu \simeq 1$. It is worth investigating this behavior more deeply. 

The proposed framework can be employed to quantify the trainability of new variational algorithms in conjunction with specific ansatz structures and optimizers. It may also assist in selecting hyperparameters that yield a better accuracy. 
Although it was shown that all the known barren-plateau-free ansatzes in the standard VQAs are implied to be classically 
simulatable~\cite{Cerezo:2023nqf}, 
numerical studies based on this framework under various practical settings may serve as a guideline for achieving quantum advantage
by circumventing this issue,
and provide insights into universal convergence behaviors, further motivating theoretical developments in VQAs.
\begin{acknowledgments}
This work was supported by the Quantum Innovation Initiative Consortium (QIIC).
\end{acknowledgments}

\appendix
\section{\label{sec:results appendix} $L$-dependence for $t=10$ and $100$}

\begin{figure*}[htbp]
    \centering
    \includegraphics[scale=0.7]{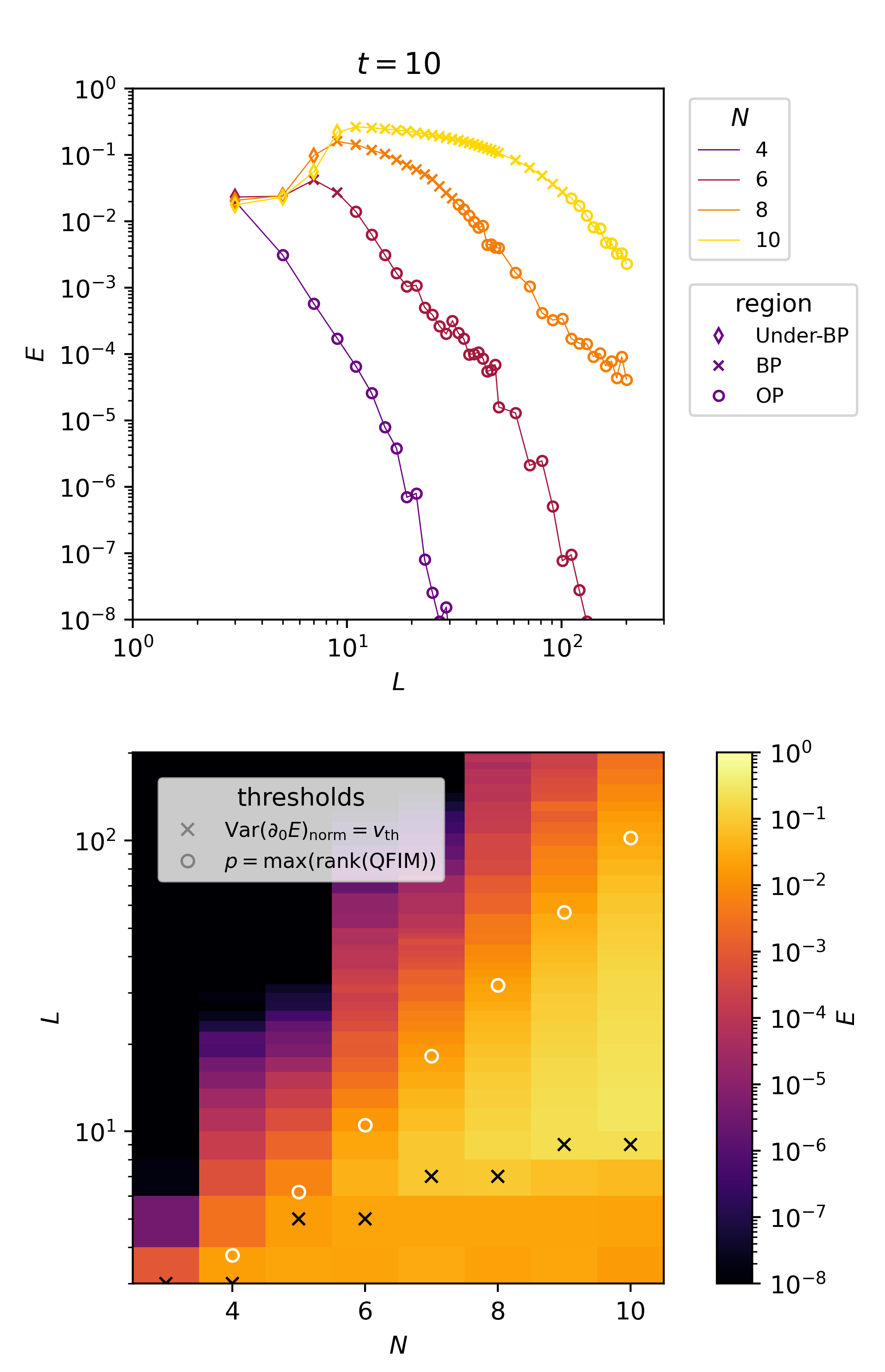}
    \includegraphics[scale=0.7]{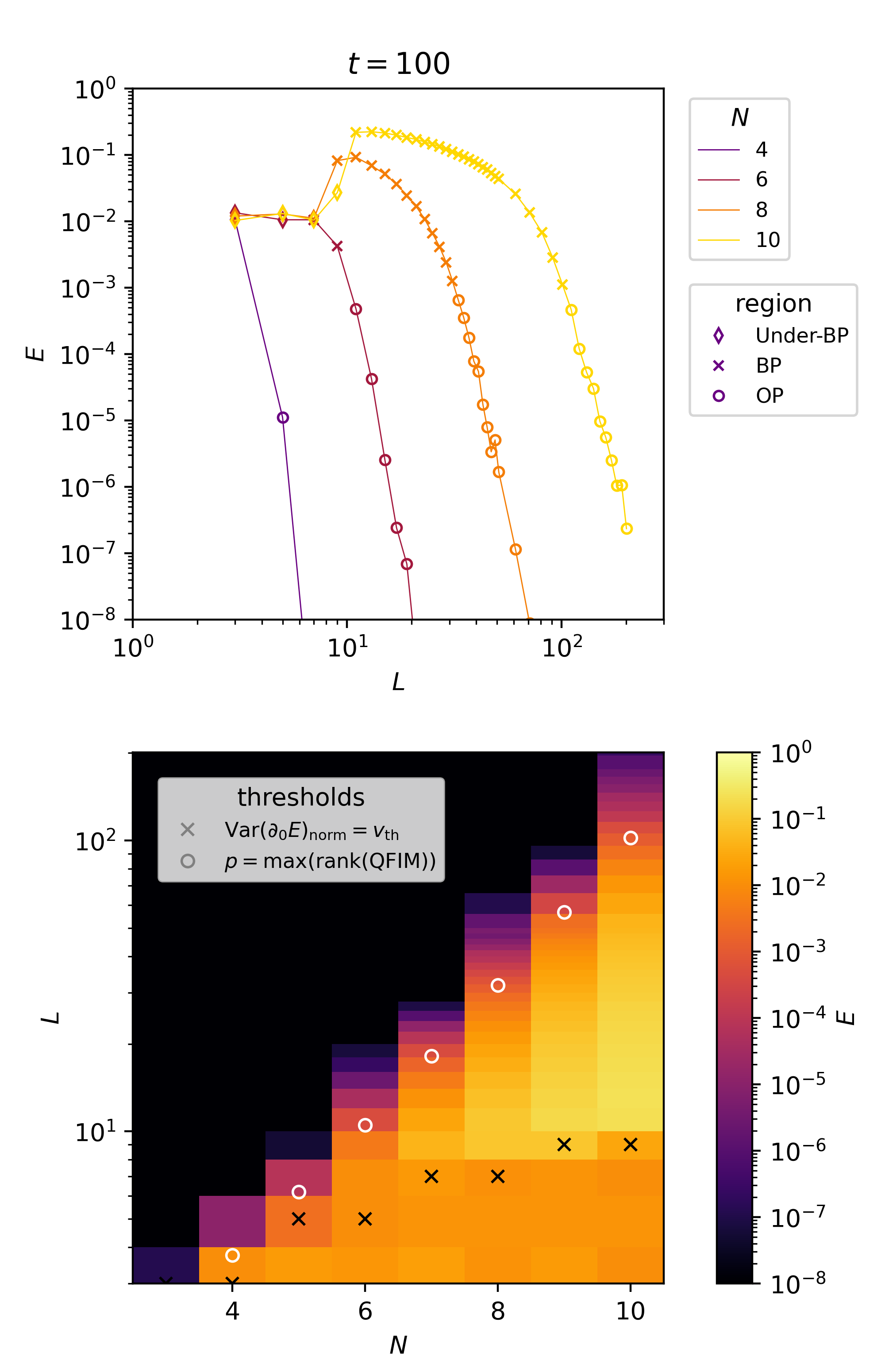}
    \caption{(Top panel) Relative residual energy $E$ as a function of $L$ at $N=4$, 6, 8 and 10 for $t=10$ (left) and $100$ (right).
    The markers show the averaged $E$ values calculated at each setting, and the lines connect the values for better visualization. (Bottom panel) Relative residual energy $E$ in $N$-$L$ plane with the white circles (black crosses) showing the OP (BP) thresholds.}
    \label{fig:LENt appendix}
\end{figure*}

Fig.~\ref{fig:LENt appendix} presents the $L$-dependence of the relative residual energy $E$ for $t=10$ and $100$, which is similar to Fig.~\ref{fig:LENt} in Section~\ref{sec:LossJump} but with fewer number of epochs~$t$.

\section{\label{sec:BP appendix}Additional results on barren plateau}

\subsection{\label{sec:BP vardiff} Variance of loss difference}

\begin{figure}[htb]
    \centering
    \includegraphics[scale=0.75]{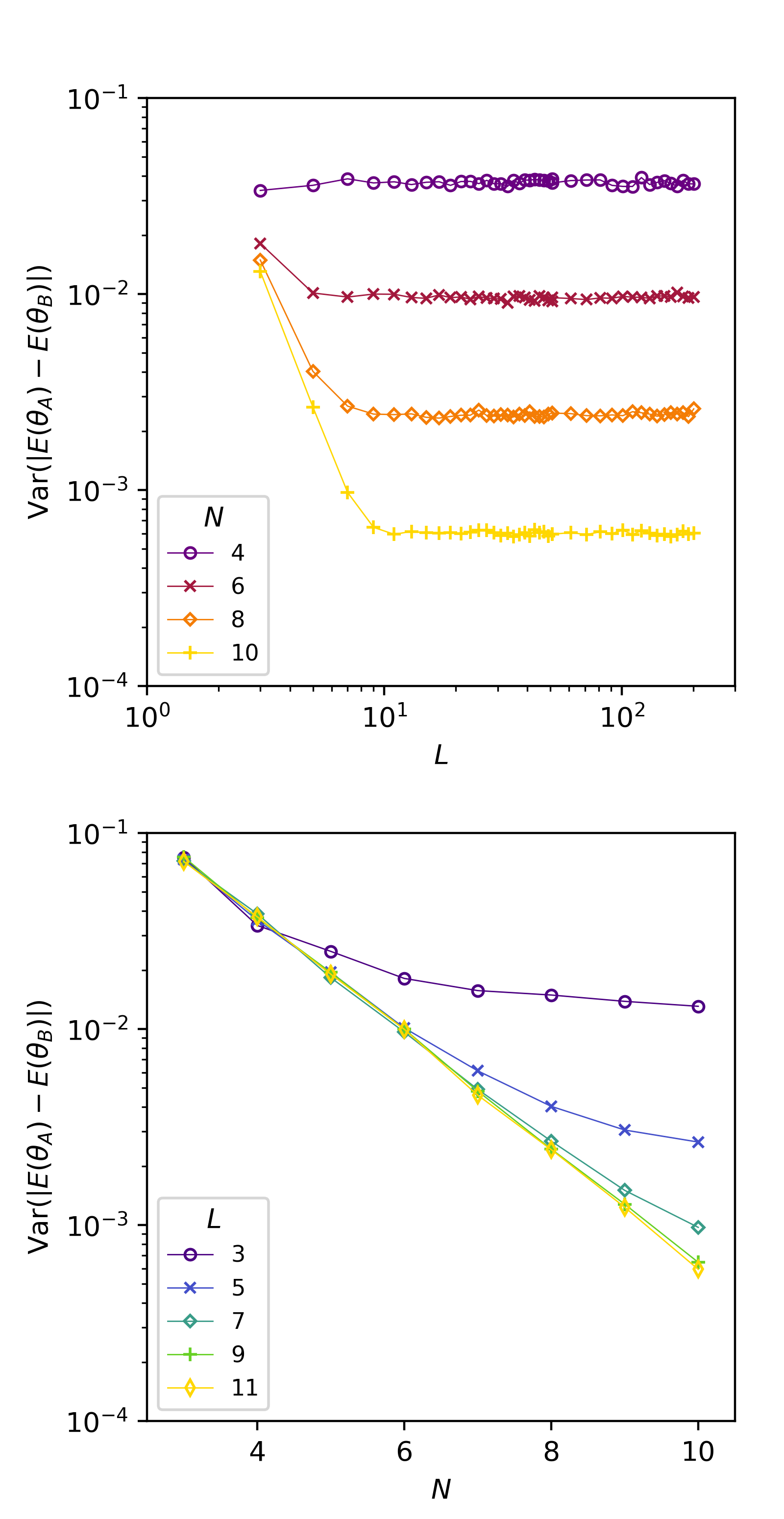}
    \caption{Variance of differences of relative residual energy $E$ between random points $\mathrm{Var}\ab(\ab|E\ab(\bm{\theta}_A) - E\ab(\bm{\theta}_B)|)$, shown as a function of $L$ ($N$) at fixed $N$ ($L$) values in the top (bottom) panel. The variance of each point is calculated from 10000 random parameter set pairs. The lines connect the calculated values for better visualization.}
    \label{fig:vardiff}
\end{figure}

Fig.~\ref{fig:vardiff} shows the behaviors of~$\mathrm{Var}\ab(\ab|E\ab(\bm{\theta}_A) - E\ab(\bm{\theta}_B)|)$. The top (bottom) panel presents the $L$ ($N$)-dependence with fixed~$N$ ($L$). One can observe from both panels that for a sufficiently deep HEA, the value of~$\mathrm{Var}\ab(\ab|E\ab(\bm{\theta}_A) - E\ab(\bm{\theta}_B)|)$ decreases exponentially with increasing value of~$N$. This is consistent with what we observed for $\mathrm{Var}(\partial_0 E)$ of partial gradients of $E$ in Fig.~\ref{fig:vargrad}.

\subsection{\label{BP frame potential}Frame potential}

\begin{figure}[htb]
    \centering
    \includegraphics[scale=0.75]{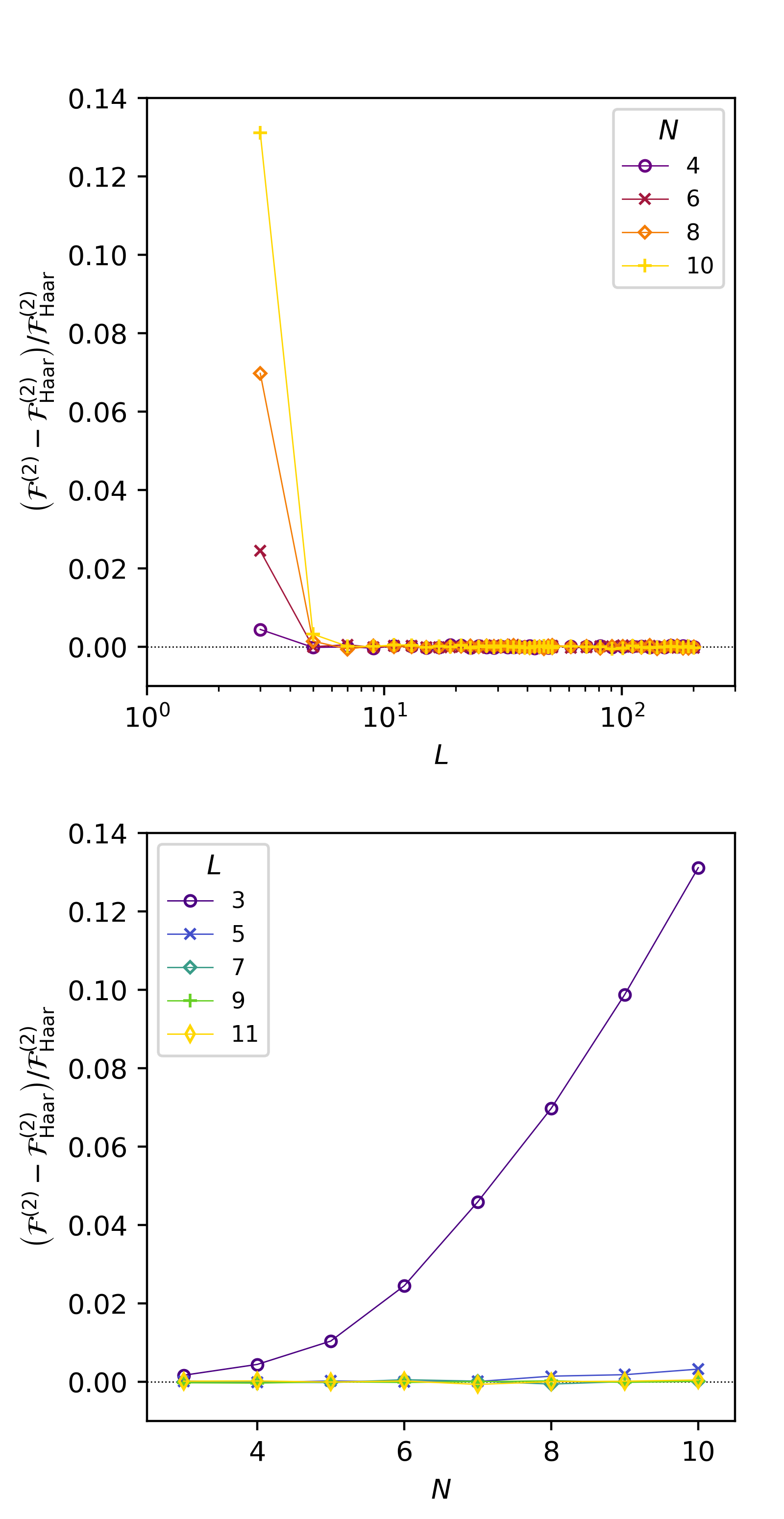}
    \caption{Relative deviation of 2nd-order general frame potential $( \mathcal{F}^{(2)} - \mathcal{F}^{(2)}_\mathrm{Haar} ) / \mathcal{F}^{(2)}_\mathrm{Haar}$, shown as a function of $L$ ($N$) at fixed $N$ ($L$) values in the top (bottom) panel. The Monte Carlo double integration is done with pairs of random $10000$ points and independent random $10000$ points. The lines connect the calculated values for better visualization.}
    \label{fig:framepotential}
\end{figure}

In this section, we discuss the expressibility of an ansatz. Here we utilize the second-order frame potential~\cite{Roberts:2016hpo} as a metric of the expressibility, according to Ref.~\cite{sim2019expressibility,Nakaji:2021uga,Holmes:2021qjw}.
It has been shown 
that the second-order frame potential~$\mathcal{F}^{(2)}$ is minimized when an ensemble of ansatz becomes a two-design, and the minimal value is given by the frame potential for the Haar random distribution denoted by~$\mathcal{F}^{(2)}_\mathrm{Haar}$~\cite{bengtsson2007geometry,renes2004symmetric,klappenecker2005mutually}. For this reason, we normalized the frame potential by the value of Haar random limit as $( \mathcal{F}^{(2)} - \mathcal{F}^{(2)}_\mathrm{Haar} ) / \mathcal{F}^{(2)}_\mathrm{Haar}$.
In Fig.~\ref{fig:framepotential}, we show the normalized frame potentials for each fixed~$(N, L)$. 
Each value of~$\mathcal{F}^{(2)}$ is approximated by the Monte Carlo integration with $10^4 \times 10^4$ random samples.
One can see that~$\mathcal{F}^{(2)}$ approaches~$\mathcal{F}^{(2)}_\mathrm{Haar}$ with increasing~$L$ and converges around $L\simeq \mathcal{O}(10)$, which is the same order as the point where $\mathrm{Var}(\partial_0 E)$ converges, as shown in Fig.~\ref{fig:vargrad}.
This result is consistent with the previous works~\cite{Nakaji:2021uga,Holmes:2021qjw}.

\section{\label{sec:op appendix}Additional results for overparametrizaiton}
\subsection{\label{sec:op exp}Exponential convergence}

\begin{figure*}[htb]
    \centering
    \includegraphics[scale=0.7]{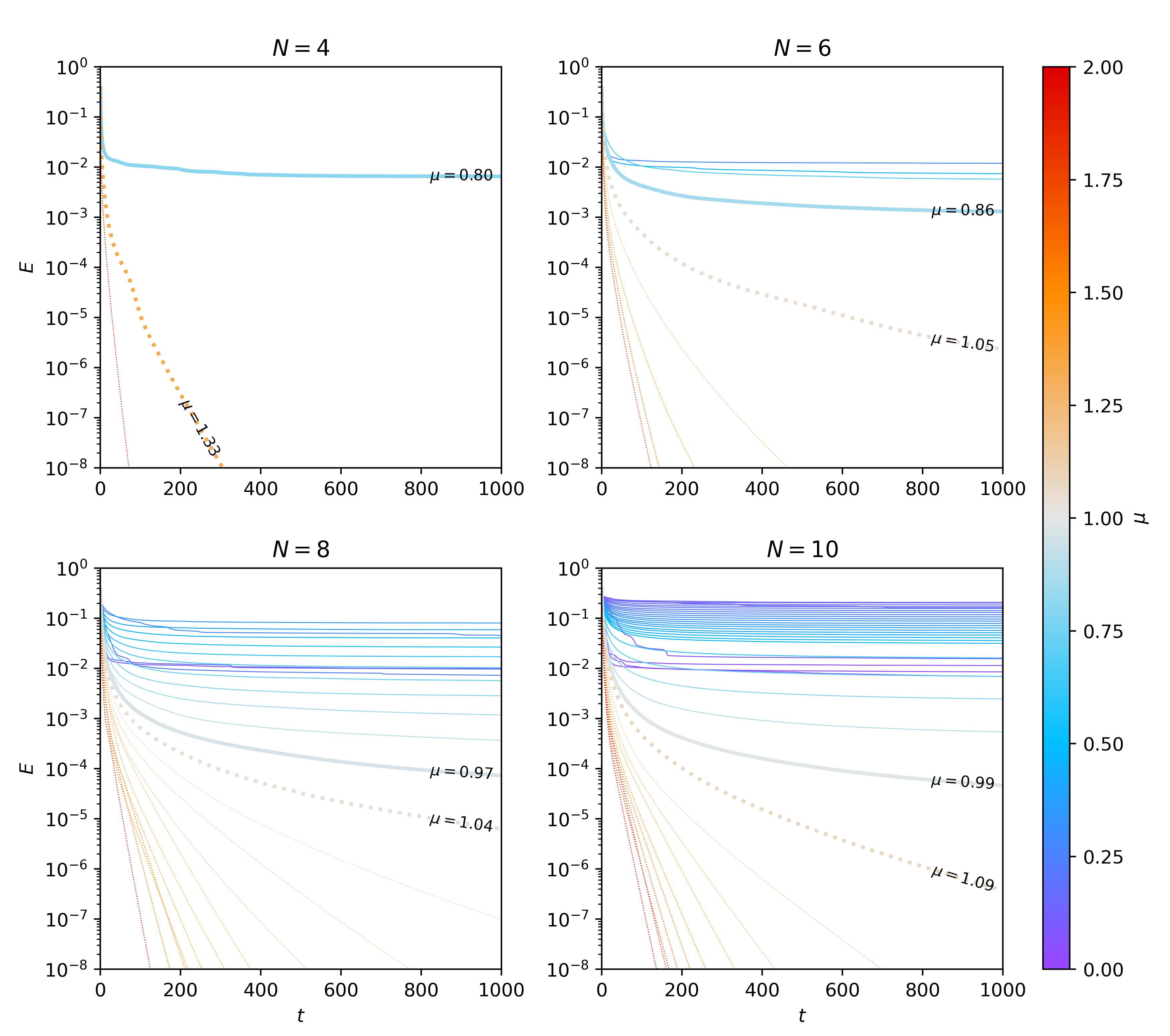}
    \caption{Relative residual energy $E$ as a function of $t$ (same results as Fig.~\ref{fig:tELN}) shown in semi-logarithmic graphs.}
    \label{fig:tELN_semilog}
\end{figure*}

To visualize exponential convergence in the overparametrization regime, we show the results of Fig.~\ref{fig:tELN}
as semi-logarithmic graphs in Fig.~\ref{fig:tELN_semilog}.
The straight lines in semi-logarithmic scale indicate exponential convergence. Such straight lines appear when the number of parameters $p = 2NL$ is sufficiently large, and it is qualitatively consistent with the previous studies.

\section{\label{sec:ERNFT vs others}Comparison of different optimization methods}

\begin{figure*}[tb]
    \centering
    \includegraphics[scale=0.7]{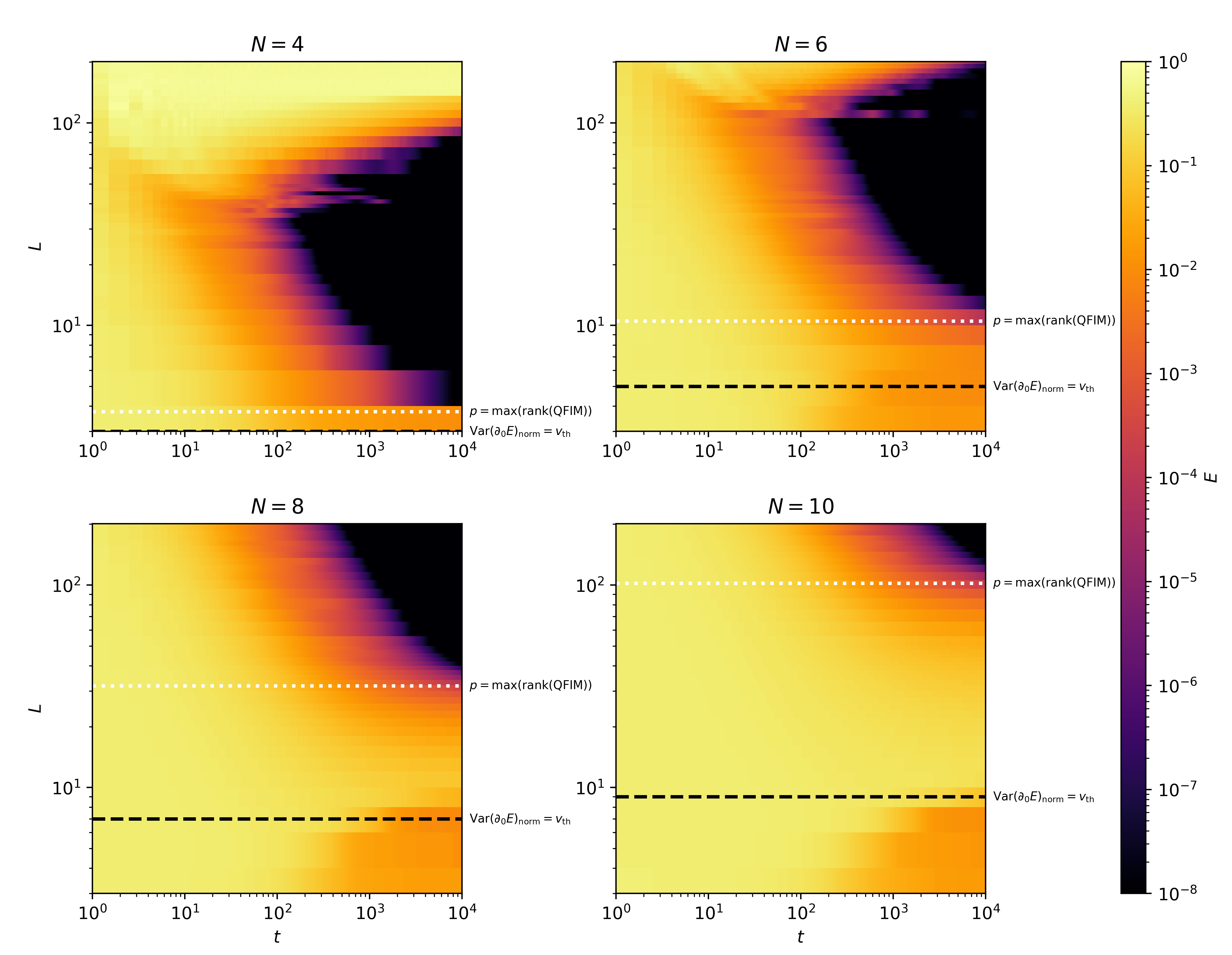}
    \caption{Heat maps of relative residual energy $E$ in the $t$-$L$ plane at $N=4$, 6, 8 and 10 obtained with the GD optimizer. Similarly to Fig.~\ref{fig:tLEmap} for the ERNFT optimizer, the $E$ value averaged over 30 runs is shown at each $(t, L)$ grid, and the black dashed and white dotted lines represent the locations where ${\mathrm{Var}(\partial_0 E)}_{\mathrm{norm}} = v_\mathrm{th}$ and the $L$ values where $p = \max(\mathrm{rank}(\mathrm{QFIM}))$, respectively.}
    \label{fig:tLEmap GD}
\end{figure*}

\begin{figure*}[tb]
    \centering
    \includegraphics[scale=0.7]{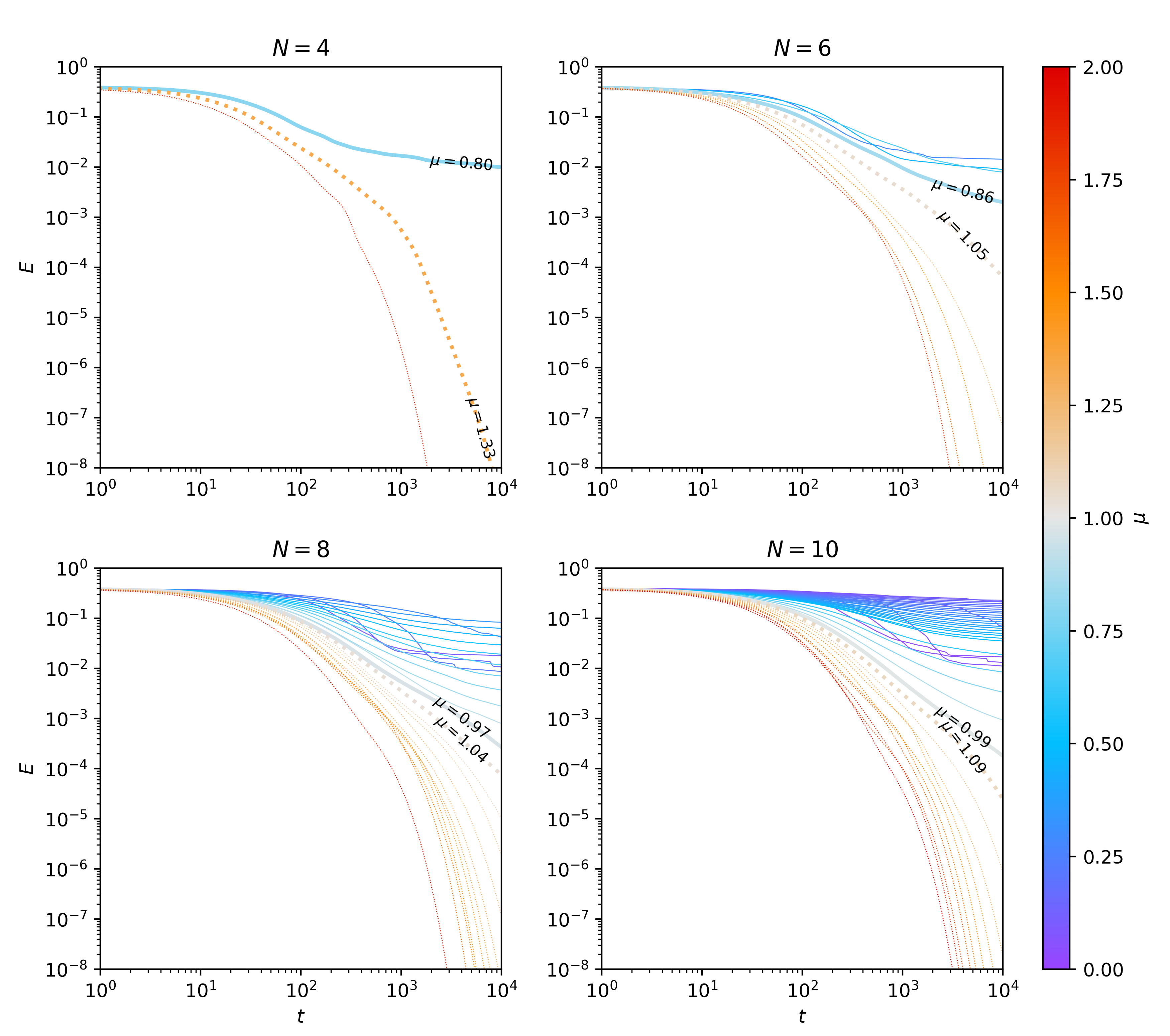}
    \caption{Relative residual energy $E$ as a function of $t$ at $N=4$, 6, 8 and 10 obtained with the GD optimizer. In each figure, the solid blue (dashed red) lines represent $E$ at $\mu < 1$ ($\mu > 1$). The residual energy for the $\mu$ value closest to 1 is shown by the gray thicker line for $N=8$ and 10, similarly to Fig.~\ref{fig:tELN} for the ERNFT optimizer.}
    \label{fig:tELN GD}
\end{figure*}

\begin{figure}[tb]
    \centering
    \includegraphics[scale=0.7]{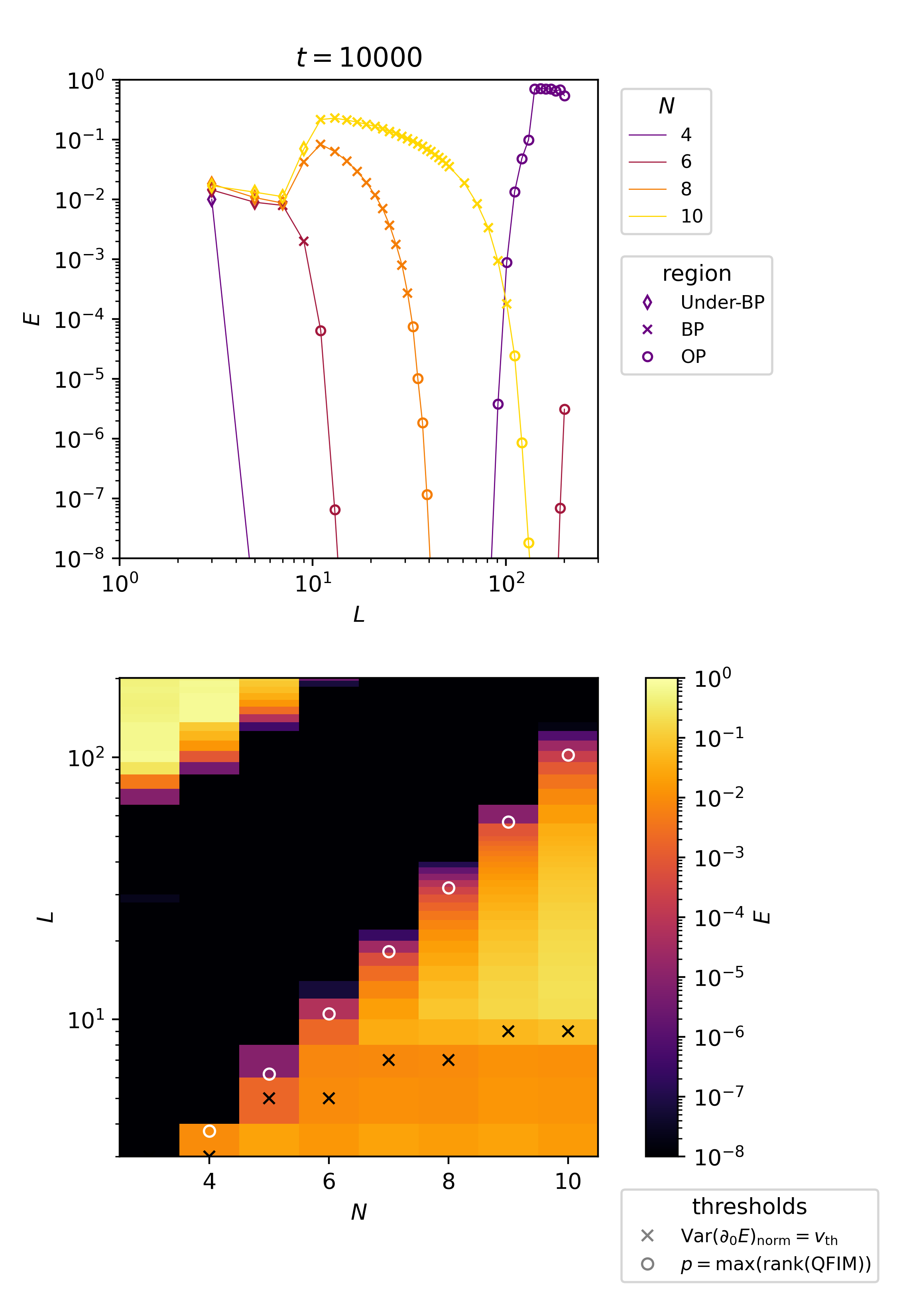}
    \caption{(Top panel) Relative residual energy $E$ as a function of $L$ at $N=4$, 6, 8 and 10 and $t=10000$ obtained with the GD optimizer. The markers show the averaged $E$ values calculated at each setting, and the lines connect the values for better visualization. (Bottom panel) Relative residual energy $E$ in $N$-$L$ plane with the white circles (black crosses) showing the OP (BP) thresholds, similarly to Fig.~\ref{fig:LENt} for the ERNFT optimizer.}
    \label{fig:LENt GD}
\end{figure}

The main results in the paper were obtained using ERNFT. Here we investigate the optimizer dependence by changing the classical optimizer while keeping the same Hamiltonian and ansatz as those used in the main text. As an alternative optimizer, we use a gradient descent (GD) optimizer,
which is widely used in the context of VQAs, often along with parameter-shift rules~\cite{PhysRevA.98.032309,PhysRevLett.118.150503}

Fig.~\ref{fig:tLEmap GD} shows the results obtained with the GD optimizer, presented similarly to Fig.~\ref{fig:tLEmap} as the dependence of relative residual energy $E$ on $L$ and $t$ at~$N=4, 6, 8$ and~$10$. The GD optimizer varies all the parameters at once in the optimization process, so the $t$-dependence shows how the $E$ varies in each update of the entire parameter set, as in Fig.~\ref{fig:tLEmap} for the ERNFT. First, one can see the three distinctive regions of (I), (II) and (III) at a sufficiently large number of $t$, and that their boundaries are still consistent with the black and white lines (same as those used in Fig.~\ref{fig:tLEmap}). This indicates that the energy convergence and its transition from the pre-BP through BP and OP regions at large $t$ are similar for the GD and ERNFT optimizers. A notable difference from Fig.~\ref{fig:tLEmap} is the ``reappearance" of an area with large $E$ values at $N=4$ and $6$ when the $L$ increases further in the OP region. We observe that this large-$E$ area moves upwards and appears at even larger $L$ values by decreasing the learning rate of the GD optimizer. From this observation, we consider this reappearance to be more relevant for the tuning of GD optimization, rather than the optimizer-independent BP or OP phenomena, but a more thorough understanding is left for the future work.

Fig.~\ref{fig:tELN GD} shows the $t$-dependence (horizontal scan) of the relative residual energy $E$ with the GD optimizer, that can be compared with Fig.~\ref{fig:tELN} for the ERNFT optimizer. At $t \simeq 100$ or higher, the $E$ exhibits a linear behavior, i.e., the power-law decay in $t$ near the QFIM boundary of~$\mu=1$, consistent with what we observe for the ERNFT optimizer. 

The relative residual energy $E$ versus $L$ in 1D and ($N$, $L$) in 2D at large $t=10000$ are shown in Fig.~\ref{fig:LENt GD}, which can be compared with the ERNFT results in Fig.~\ref{fig:LENt}. They look very similar for $N=8$ and $10$, including the locations of jumps to BP region and subsequent exponential falling in the OP region, while at a smaller $N$ there is a distinctive difference due to the reappearance of the area with large $E$ values mentioned above. The boundaries of the BP and OP regions observed before the reappearance are quite consistent with the calculations of the maximum QFIM rank and gradient thresholds.

\bibliography{bibliography/qc-yokogawa_aps}

@article{Larocca:2024plh,
    author = {Larocca, Martin and Thanasilp, Supanut and Wang, Samson and Sharma, Kunal and Biamonte, Jacob and Coles, Patrick J. and Cincio, Lukasz and McClean, Jarrod R. and Holmes, Zo{\"e} and Cerezo, M.},
    title = "{Barren plateaus in variational quantum computing}",
    eprint = "2405.00781",
    archivePrefix = "arXiv",
    primaryClass = "quant-ph",
    reportNumber = "LA-UR-24-23934",
    doi = "10.1038/s42254-025-00813-9",
    journal = "Nature Rev. Phys.",
    volume = "7",
    number = "4",
    pages = "174--189",
    year = "2025"
}

@article{McClean:2018jps,
    author = "McClean, Jarrod R. and Boixo, Sergio and Smelyanskiy, Vadim N. and Babbush, Ryan and Neven, Hartmut",
    title = "{Barren plateaus in quantum neural network training landscapes}",
    eprint = "1803.11173",
    archivePrefix = "arXiv",
    primaryClass = "quant-ph",
    doi = "10.1038/s41467-018-07090-4",
    journal = "Nature Commun.",
    volume = "9",
    pages = "4812",
    year = "2018"
}

@article{miao2024equivalence,
  title={Equivalence of cost concentration and gradient vanishing for quantum circuits: an elementary proof in the Riemannian formulation},
  author={Miao, Qiang and Barthel, Thomas},
  journal={Quantum Science and Technology},
  volume={9},
  number={4},
  pages={045039},
  year={2024},
  publisher={IOP Publishing}
}

@article{arrasmith2022equivalence,
  title={Equivalence of quantum barren plateaus to cost concentration and narrow gorges},
  author={Arrasmith, Andrew and Holmes, Zo{\"e} and Cerezo, Marco and Coles, Patrick J},
  journal={Quantum Science and Technology},
  volume={7},
  number={4},
  pages={045015},
  year={2022},
  publisher={IOP Publishing}
}

@article{Thanasilp:2021axb,
    author = "Thanasilp, Supanut and Wang, Samson and Nghiem, Nhat Anh and Coles, Patrick J. and Cerezo, Marco",
    title = "{Subtleties in the trainability of quantum machine learning models}",
    eprint = "2110.14753",
    archivePrefix = "arXiv",
    primaryClass = "quant-ph",
    reportNumber = "LA-UR-21-30290",
    doi = "10.1007/s42484-023-00103-6",
    journal = "Quantum Machine Intelligence",
    volume = "5",
    number = "1",
    pages = "21",
    year = "2023"
}

@article{Nakanishi:2019rrm,
    author = "Nakanishi, Ken M. and Fujii, Keisuke and Todo, Synge",
    title = "{Sequential minimal optimization for quantum-classical hybrid algorithms}",
    eprint = "1903.12166",
    archivePrefix = "arXiv",
    primaryClass = "quant-ph",
    doi = "10.1103/physrevresearch.2.043158",
    journal = "Phys. Rev. Res.",
    volume = "2",
    number = "4",
    pages = "043158",
    year = "2020"
}

@article{Holmes:2021qjw,
    author = {Holmes, Zo{\"e} and Sharma, Kunal and Cerezo, M. and Coles, Patrick J.},
    title = "{Connecting Ansatz Expressibility to Gradient Magnitudes and Barren Plateaus}",
    eprint = "2101.02138",
    archivePrefix = "arXiv",
    primaryClass = "quant-ph",
    reportNumber = "LA-UR-21-20034",
    doi = "10.1103/PRXQuantum.3.010313",
    journal = "PRX Quantum",
    volume = "3",
    number = "1",
    pages = "010313",
    year = "2022"
}

@article{Larocca:2021jub,
    author = "Larocca, Mart{\'\i}n and Ju, Nathan and Garc{\'\i}a-Mart{\'\i}n, Diego and Coles, Patrick J. and Cerezo, Marco",
    title = "{Theory of overparametrization in quantum neural networks}",
    eprint = "2109.11676",
    archivePrefix = "arXiv",
    primaryClass = "quant-ph",
    reportNumber = "LA-UR-21-29233",
    doi = "10.1038/s43588-023-00467-6",
    journal = "Nature Computat. Sci.",
    volume = "3",
    number = "6",
    pages = "542--551",
    year = "2023"
}

@article{Cerezo:2023nqf,
    author = "Cerezo, M. and others",
    title = "{Does provable absence of barren plateaus imply classical simulability?}",
    eprint = "2312.09121",
    archivePrefix = "arXiv",
    primaryClass = "quant-ph",
    reportNumber = "LA-UR-23-33705",
    doi = "10.1038/s41467-025-63099-6",
    journal = "Nature Commun.",
    volume = "16",
    number = "1",
    pages = "7907",
    year = "2025"
}

@article{Liu:2022eqa,
    author = "Liu, Junyu and Najafi, Khadijeh and Sharma, Kunal and Tacchino, Francesco and Jiang, Liang and Mezzacapo, Antonio",
    title = "{Analytic Theory for the Dynamics of Wide Quantum Neural Networks}",
    eprint = "2203.16711",
    archivePrefix = "arXiv",
    primaryClass = "quant-ph",
    doi = "10.1103/PhysRevLett.130.150601",
    journal = "Phys. Rev. Lett.",
    volume = "130",
    number = "15",
    pages = "150601",
    year = "2023"
}

@article{Suzuki:2020nbf,
    author = "Suzuki, Yasunari and others",
    title = "{Qulacs: a fast and versatile quantum circuit simulator for research purpose}",
    eprint = "2011.13524",
    archivePrefix = "arXiv",
    primaryClass = "quant-ph",
    doi = "10.22331/q-2021-10-06-559",
    journal = "Quantum",
    volume = "5",
    pages = "559",
    year = "2021"
}

@misc{qiskit2024,
title = {Quantum computing with {Q}iskit},
author = {Javadi-Abhari, Ali and Treinish, Matthew and Krsulich, Kevin and Wood, Christopher J. and Lishman, Jake and Gacon, Julien and Martiel, Simon and Nation, Paul D. and Bishop, Lev S. and Cross, Andrew W. and Johnson, Blake R. and Gambetta, Jay M.},
year = {2024},
doi = {10.48550/arXiv.2405.08810},
eprint = {2405.08810},
archivePrefix = {arXiv},
primaryClass = {quant-ph},
}

@article{Kandala:2017vok,
    author = "Kandala, Abhinav and Mezzacapo, Antonio and Temme, Kristan and Takita, Maika and Brink, Markus and Chow, Jerry M. and Gambetta, Jay M.",
    title = "{Hardware-efficient variational quantum eigensolver for small molecules and quantum magnets}",
    eprint = "1704.05018",
    archivePrefix = "arXiv",
    primaryClass = "quant-ph",
    doi = "10.1038/nature23879",
    journal = "Nature",
    volume = "549",
    number = "7671",
    pages = "242--246",
    year = "2017"
}

@inproceedings{klappenecker2005mutually,
  title={Mutually unbiased bases are complex projective 2-designs},
  author={Klappenecker, Andreas and Rotteler, Martin},
  booktitle={Proceedings. International Symposium on Information Theory, 2005. ISIT 2005.},
  pages={1740--1744},
  year={2005},
  organization={IEEE}
}

@article{renes2004symmetric,
  title={Symmetric informationally complete quantum measurements},
  author={Renes, Joseph M and Blume-Kohout, Robin and Scott, Andrew J and Caves, Carlton M},
  journal={Journal of Mathematical Physics},
  volume={45},
  number={6},
  pages={2171--2180},
  year={2004},
  publisher={American Institute of Physics}
}

@article{Roberts:2016hpo,
    author = "Roberts, Daniel A. and Yoshida, Beni",
    title = "{Chaos and complexity by design}",
    eprint = "1610.04903",
    archivePrefix = "arXiv",
    primaryClass = "quant-ph",
    doi = "10.1007/JHEP04(2017)121",
    journal = "JHEP",
    volume = "04",
    pages = "121",
    year = "2017"
}

@article{sim2019expressibility,
  title={Expressibility and entangling capability of parameterized quantum circuits for hybrid quantum-classical algorithms},
  author={Sim, Sukin and Johnson, Peter D and Aspuru-Guzik, Al{\'a}n},
  journal={Advanced Quantum Technologies},
  volume={2},
  number={12},
  pages={1900070},
  year={2019},
  publisher={Wiley Online Library}
}

@book{bengtsson2007geometry,
  title={Geometry of Quantum States: An Introduction to Quantum Entanglement},
  author={Bengtsson, I. and Zyczkowski, K.},
  isbn={9781139453462},
  url={https://books.google.co.jp/books?id=aA4vXMbuOTUC},
  year={2007},
  publisher={Cambridge University Press}
}

@article{Nakaji:2021uga,
    author = "Nakaji, Kouhei and Yamamoto, Naoki",
    title = "{Expressibility of the alternating layered ansatz for quantum computation}",
    eprint = "2005.12537",
    archivePrefix = "arXiv",
    primaryClass = "quant-ph",
    doi = "10.22331/q-2021-04-19-434",
    journal = "Quantum",
    volume = "5",
    pages = "434",
    year = "2021"
}

@incollection{Petz:2010avh,
  title={Introduction to quantum Fisher information},
  author={Petz, D{\'e}nes and Ghinea, Catalin},
  booktitle={Quantum probability and related topics},
  pages={261--281},
  year={2011},
  publisher={World Scientific}
}

@article{Wiersema:2020ipa,
    author = "Wiersema, Roeland and Zhou, Cunlu and de Sereville, Yvette and Carrasquilla, Juan Felipe and Kim, Yong Baek and Yuen, Henry",
    title = "{Exploring Entanglement and Optimization within the Hamiltonian Variational Ansatz}",
    eprint = "2008.02941",
    archivePrefix = "arXiv",
    primaryClass = "quant-ph",
    doi = "10.1103/PRXQuantum.1.020319",
    journal = "PRX Quantum",
    volume = "1",
    number = "2",
    pages = "020319",
    year = "2020"
}

@article{Larocca:2021ksf,
    author = "Larocca, Martin and Czarnik, Piotr and Sharma, Kunal and Muraleedharan, Gopikrishnan and Coles, Patrick J. and Cerezo, M.",
    title = "{Diagnosing Barren Plateaus with Tools from Quantum Optimal Control}",
    eprint = "2105.14377",
    archivePrefix = "arXiv",
    primaryClass = "quant-ph",
    reportNumber = "LA-UR-21-24973",
    doi = "10.22331/q-2022-09-29-824",
    journal = "Quantum",
    volume = "6",
    pages = "824",
    year = "2022"
}

@article{Cerezo:2020mtn,
    author = "Cerezo, M. and Sone, Akira and Volkoff, Tyler and Cincio, Lukasz and Coles, Patrick J.",
    title = "{Cost function dependent barren plateaus in shallow parametrized quantum circuits}",
    eprint = "2001.00550",
    archivePrefix = "arXiv",
    primaryClass = "quant-ph",
    reportNumber = "LA-UR-19-32681, LA-UR-19-32681",
    doi = "10.1038/s41467-021-21728-w",
    journal = "Nature Commun.",
    volume = "12",
    number = "1",
    pages = "1791",
    year = "2021"
}

@article{Marrero:2020gvt,
    title={Entanglement-induced barren plateaus},
  author={Ortiz Marrero, Carlos and Kieferov{\'a}, M{\'a}ria and Wiebe, Nathan},
  journal={PRX quantum},
  volume={2},
  number={4},
  pages={040316},
  year={2021},
  publisher={APS}
}

@article{Wang:2020yjh,
    author = "Wang, Samson and Fontana, Enrico and Cerezo, M. and Sharma, Kunal and Sone, Akira and Cincio, Lukasz and Coles, Patrick J.",
    title = "{Noise-Induced Barren Plateaus in Variational Quantum Algorithms}",
    eprint = "2007.14384",
    archivePrefix = "arXiv",
    primaryClass = "quant-ph",
    reportNumber = "LA-UR-20-25526",
    doi = "10.1038/s41467-021-27045-6",
    journal = "Nature Commun.",
    volume = "12",
    pages = "6961",
    year = "2021"
}

@article{Patti:2020ach,
    author = "Patti, Taylor L. and Najafi, Khadijeh and Gao, Xun and Yelin, Susanne F.",
    title = "{Entanglement devised barren plateau mitigation}",
    eprint = "2012.12658",
    archivePrefix = "arXiv",
    primaryClass = "quant-ph",
    doi = "10.1103/PhysRevResearch.3.033090",
    journal = "Phys. Rev. Res.",
    volume = "3",
    number = "3",
    pages = "033090",
    year = "2021"
}

@article{Kiani:2020bwb,
    title={Learning unitaries by gradient descent},
  author={Kiani, Bobak Toussi and Lloyd, Seth and Maity, Reevu},
  journal={arXiv preprint arXiv:2001.11897},
  year={2020}
}

@article{Kim:2020luc,
    author = "Kim, Joonho and Kim, Jaedeok and Rosa, Dario",
    title = "{Universal effectiveness of high-depth circuits in variational eigenproblems}",
    eprint = "2010.00157",
    archivePrefix = "arXiv",
    primaryClass = "quant-ph",
    doi = "10.1103/PhysRevResearch.3.023203",
    journal = "Phys. Rev. Res.",
    volume = "3",
    number = "2",
    pages = "023203",
    year = "2021"
}

@software{masumoto_2025_15832836,
  author       = {Masumoto, Naoyuki and
                  Matsumoto, Keita and
                  Miyaji, Kosuke and
                  Miyanaga, Takafumi and
                  Mori, Toshio and
                  Tsukano, Satoyuki and
                  Uchida, Ryo},
  title        = {QURI Parts OQTOPUS: A Library for Running Quantum
                   Computers on OQTOPUS Cloud
                  },
  month        = jul,
  year         = 2025,
  publisher    = {Zenodo},
  version      = {v1.0.3},
  doi          = {10.5281/zenodo.15832836},
  url          = {https://doi.org/10.5281/zenodo.15832836},
  swhid        = {swh:1:dir:9870222aa4b86b2a5687d9f2b31e4e86b2c98204
                   ;origin=https://doi.org/10.5281/zenodo.14974331;vi
                   sit=swh:1:snp:520237ec53278b456edfbcd189dcd9d14963
                   1139;anchor=swh:1:rel:ff01a1a7f0ab773bfe6bc5582b7a
                   7bcdc88c6555;path=oqtopus-team-quri-parts-
                   oqtopus-90714a5
                  },
}

@article{nicoli2023physics,
  title={Physics-informed bayesian optimization of variational quantum circuits},
  author={Nicoli, Kim and Anders, Christopher J and Funcke, Lena and Hartung, Tobias and Jansen, Karl and K{\"u}hn, Stefan and M{\"u}ller, Klaus-Robert and Stornati, Paolo and Kessel, Pan and Nakajima, Shinichi},
  journal={Advances in Neural Information Processing Systems},
  volume={36},
  pages={18341--18376},
  year={2023}
}

@article{PhysRevA.98.032309,
  title = {Quantum circuit learning},
  author = {Mitarai, K. and Negoro, M. and Kitagawa, M. and Fujii, K.},
  journal = {Phys. Rev. A},
  volume = {98},
  issue = {3},
  pages = {032309},
  numpages = {6},
  year = {2018},
  month = {Sep},
  publisher = {American Physical Society},
  doi = {10.1103/PhysRevA.98.032309},
  url = {https://link.aps.org/doi/10.1103/PhysRevA.98.032309}
}

@article{PhysRevLett.118.150503,
  title = {Hybrid Quantum-Classical Approach to Quantum Optimal Control},
  author = {Li, Jun and Yang, Xiaodong and Peng, Xinhua and Sun, Chang-Pu},
  journal = {Phys. Rev. Lett.},
  volume = {118},
  issue = {15},
  pages = {150503},
  numpages = {5},
  year = {2017},
  month = {Apr},
  publisher = {American Physical Society},
  doi = {10.1103/PhysRevLett.118.150503},
  url = {https://link.aps.org/doi/10.1103/PhysRevLett.118.150503}
}

@article{Zhang:2024ktz,
    author = "Zhang, Bingzhi and Liu, Junyu and Jiang, Liang and Zhuang, Quntao",
    title = "{Quantum-data-driven dynamical transition in quantum learning}",
    eprint = "2410.01955",
    archivePrefix = "arXiv",
    primaryClass = "quant-ph",
    doi = "10.1038/s41534-025-01079-w",
    journal = "npj Quantum Inf.",
    volume = "11",
    number = "1",
    pages = "132",
    year = "2025"
}

@inproceedings{you2023analyzing,
  title={Analyzing convergence in quantum neural networks: deviations from neural tangent kernels},
  author={You, Xuchen and Chakrabarti, Shouvanik and Chen, Boyang and Wu, Xiaodi},
  booktitle={International Conference on Machine Learning},
  pages={40199--40224},
  year={2023},
  organization={PMLR}
}

\end{document}